\documentclass[prd,aps,amsfonts,showpacs,longbibliography,twocolumn,notitlepage,superscriptaddress,nofootinbib]{revtex4-1}
\usepackage[dvipsnames]{xcolor}
\usepackage{multirow}
\usepackage{tabularx}
\usepackage{amsmath}
\usepackage{graphicx}
\usepackage{comment}

\usepackage[colorlinks=true, urlcolor=violet, linkcolor=blue, citecolor=red, hyperindex=true, linktocpage=true]{hyperref}

\usepackage{slashed}

\allowdisplaybreaks

\numberwithin{thm}{section}

\newcommand{\nn}{\nonumber\\}

\renewcommand{\thesubsection}{\thesection.\arabic{subsection}}

\makeatletter
\renewcommand{\p@subsection}{}
\renewcommand{\p@subsubsection}{}
\makeatother

\usepackage{xcolor}
\usepackage{mathtools}
\usepackage{ragged2e}

\newcommand{\andy}[1]{{\color{red}[andy: \textsf{#1}]}}

\newcommand{\fedor}[1]{{\color{orange}[fedor: \textsf{#1}]}}

\newcommand{\ud}{\mathrm{d}}

\newcommand\cA{\mathcal{A}}

\newcommand\cF{\mathcal{F}}

\newcommand{\ii}{\mathrm{i}}

\newcommand{\tr}{\mathrm{tr}}

\newcommand{\p}{\partial}

\newcommand{\abs}[1]{\lvert{#1}\rvert}

\makeatletter 
    
\renewcommand\onecolumngrid{% <<<<<<
\do@columngrid{one}{\@ne}%
\def\set@footnotewidth{\onecolumngrid}% <<<<<<<<<<<<<<<<
\def\footnoterule{\kern-6pt\hrule width 1.5in\kern6pt}%
}

\renewcommand\twocolumngrid{% <<<<<<
        \def\footnoterule{% restore rule
        \dimen@\skip\footins\divide\dimen@\thr@@
        \kern-\dimen@\hrule width.5in\kern\dimen@}
        \do@columngrid{mlt}{\tw@}
}%

\makeatother   

\usepackage{dsfont}

\begin{document}

\title{Entropic Order}

\author{Yiqiu Han}
\affiliation{Department of Physics and Center for Theory of Quantum Matter, University of Colorado, Boulder, CO 80309, USA}

\author{Xiaoyang Huang}
\affiliation{Department of Physics and Center for Theory of Quantum Matter, University of Colorado, Boulder, CO 80309, USA}

\author{Zohar Komargodski}
\affiliation{Simons Center for Geometry and Physics, SUNY, Stony Brook, NY 11794, USA}

\author{Andrew Lucas}
\email{andrew.j.lucas@colorado.edu}
\affiliation{Department of Physics and Center for Theory of Quantum Matter, University of Colorado, Boulder, CO 80309, USA}

\author{Fedor K. Popov}
\affiliation{Simons Center for Geometry and Physics, SUNY, Stony Brook, NY 11794, USA}
\date{\today}

\begin{abstract}
Ordered phases of matter, such as solids, ferromagnets, superfluids, or quantum topological order, typically only exist at low temperatures. Despite this conventional wisdom, we present explicit local models in which all such phases  persist to  arbitrarily high temperature.   This is possible since %analogous to the Pomeranchuk effect in $^3\mathrm{He}$: 
order in one degree of freedom can enable other degrees of freedom to strongly fluctuate, leading to ``entropic order", whereby typical high energy states are ordered.
Our construction, which utilizes interacting bosons, avoids existing no-go theorems on long-range order or entanglement at high temperature.    We propose a simple model  for high-temperature superconductivity using these general principles. 
%Using our framework, we also present a model with superconductivity at very high temperatures.

%\andy{Interacting bosons can lead to entropic order at any temperature, avoiding existing no-go theorems on long-range order or entanglement at high temperature.   These counterexamples to the conventional wisdom of statistical physics may lead to new strategies to design ordered phases, such as superconductors, at very high temperature.}
%Using our framework, we also present a model with superconductivity at very high temperatures.

\end{abstract}

\maketitle

\emph{Introduction.---} Statistical mechanics is the theory of simple collective phenomena that arise out of many-body physical systems.  The most striking phenomenon is that of a phase transition: as a parameter -- which we take here to be temperature $T$ -- is tuned beyond a critical value $T_{\mathrm{c}}$, the macroscopic phase of matter abruptly changes.   For example, when solid ice is heated above $T_{\mathrm{c}} = 0^\circ$~C, it abruptly melts into liquid water.  
%When ferromagnetic iron is raised to sufficiently high temperature, the spontaneous magnetization of the material is lost and it becomes paramagnetic.

Here, the low-temperature phase is more ordered than the high-temperature phase.   Solid ice forms a crystal and spontaneously breaks translation and rotation symmetries: the crystalline lattice has a specific orientation and the atoms prefer to be in specific places relative to others,  over arbitrarily long length scales.   In liquid water, there is no long-range order: translation and rotation symmetry are restored.  Breaking these symmetries make ice an \emph{ordered phase} and water a \emph{disordered phase}.   

In thermodynamics, systems minimize the free energy \begin{equation}
    F = E-TS, \label{eq:FETS}
\end{equation}
where $E$ denotes energy, $T$ temperature, and $S$ entropy.
If we cross a phase transition from A to B at some temperature $T_\mathrm{c}$ (in this paper, other external parameters, e.g. volume, are fixed), the high-temperature phase has higher entropy at $T>T_\mathrm{c}$.  This is readily seen from the laws of thermodynamics, which imply that $F=\min(F_{\mathrm{A}},F_{\mathrm{B}})$ and \begin{equation}
    S = -\frac{\partial F}{\partial T}.
\end{equation}
Entropy is usually associated with ``disorder", so we  expect the disordered phase at high $T$.

But sometimes, the ordered phase is at high temperature. This is  because the ordered phase can have higher entropy, however unlikely it sounds. We  will say that such a phase has \emph{entropic order}.  Experimental realizations include crystalline ordering in the Rochelle salt \cite{kao2004dielectric} and inverse melting \cite{plazanet, leo2001}.   But the most famous example is the Pomeranchuk effect in $^3\mathrm{He}$: at $T< 10^{-7}$ K (and 30 atm pressure) we find a liquid, while for $10^{-7}$ K $< T < 1$ K we find a solid \cite{pomeranchuk1950theory}.  When the atoms lock into a crystal, atomic isospin degrees of freedom can freely fluctuate, while they cannot in the liquid.  By sacrificing the translational entropy and forming a crystal, we overcompensate with extra isospin entropy. Analogous phenomena have also been found in magic-angle graphene \cite{rozen2021entropic}.  A somewhat related mechanism at low temperatures is known as ``order by disorder" \cite{villain}.

In all of these examples, upon sufficiently heating the system, one again finds a disordered phase.  Under seemingly mild conditions, which we review in the Supplementary Material (SM), one can even prove  that the high temperature phase of any discrete lattice model must be disordered and have no quantum entanglement~\cite{kliesch2014locality,Bakshi:2024cqr}.  At the same time, certain  quantum field theories order at \emph{arbitrarily high} temperature~\cite{Chai:2020onq,Chai:2020zgq,Liendo:2022bmv,Chai:2020hnu,Chaudhuri:2020xxb,Bajc:2020gpa,Nakayama:2021fgy,Chaudhuri:2021dsq,Agrawal:2021alq,Hawashin:2024dpp,Komargodski:2024zmt} 
(Closely related  constructions using the AdS/CFT correspondence appeared in~\cite{Buchel:2020jfs,Buchel:2020thm,Buchel:2021ead,Buchel:2022zxl,Buchel:2023zpe}, and some nonlocal theories were considered in~\cite{Chai:2021tpt,Chai:2021djc}). 

It is an outstanding question whether or not order as $T\rightarrow\infty$ is an ``artifact" of field theory:  is such order possible in simple lattice models?  What is the physical mechanism for it?
%or even in an experimentally realizable setting?

We will answer the above  questions, presenting explicit models with order as $T\rightarrow\infty$, both in lattice models and field theory.   Illustrative examples include high-temperature ferromagnets, solids, superfluids, quantum topological order, and superconductivity. There is a unifying principle behind all of our examples: interacting bosons have unbounded number fluctuations which are enhanced by the existence of order in a second degree of freedom, e.g. spins.  The bosonic entropy overcompensates for the reduced spin entropy due to ordering, and we find that most states at fixed (high) energy are ordered.  This is why we adopt the terminology ``entropic order".  We show that this  mechanism underlies high-$T$ order in the field theory models mentioned above. We emphasize the toy models for very high temperature superconductivity, which may guide a search for this long-sought phenomenon.

%\Zohar{An important fact is that there is a unifying principle behind all of our examples of high temperature order. 
%The principle is that in the presence of multiple degrees of freedom, it is often advantageous to order a few of them, since this might lead to many micro-states for the remaining, disordered ones, more than the number of micro-states (at fixed energy) when all the degrees of freedom are disordered. This is why we adopt the terminology ``entropic order.'' } \andy{I merged these comments in above?}

\emph{Brief review.---} Let us briefly recall basic statistical mechanics, using a classical example for illustration (details of both classical and quantum models are in the SM).  Consider a two-dimensional square lattice, where at every vertex $v$ we choose to put $n_v \in \lbrace 0,1\rbrace$ particles.  The collection of all $\mathbf{n} = \lbrace n_v\rbrace$ is called a \emph{microstate}.  In statistical mechanics, the probability of observing a particular microstate at temperature $T=1/\beta$ is \begin{equation}
    \mathbb{P}(\mathbf{n}) = \frac{\mathrm{e}^{-\beta H(\mathbf{n})}}{Z(\beta)} \label{eq:thermalensemble}
\end{equation}
where $H(\mathbf{n})$ is the energy (Hamiltonian) of the microstate, and the partition function \begin{equation}
    Z(\beta) = \sum_{\mathbf{n}} \mathrm{e}^{-\beta H(\mathbf{n})} = \mathrm{e}^{-\beta F(\beta)} \label{eq:Zdef}
\end{equation}
both normalizes the probability distribution, and defines the free energy $F$. A system is a ``many-body system" if $H(\mathbf{n})$ can be expressed as a sum of terms, each of which only depends on a finite number of $n_v$, which we take here to be nearest-neighbors on the lattice.  

%Notice that we will not restrict the sum over possible microstates $\mathbf{n}$ in \eqref{eq:Zdef}; if this is done, ordered states at $T=\infty$ are less surprising (see SM):  a hardcore lattice gas, where the only valid configurations have $n_un_v=0$ on all edges, orders if we restrict to a high-charge ensemble \cite{}.

A famous illustrative example for how phase transitions usually arise is the lattice gas \cite{yang1952}
\begin{equation}\label{IntroModel}
    H = U\sum_{u\sim v} n_u n_v - \mu\sum_v n_v,
\end{equation}
where $u\sim v$ denote nearest neighbors on the lattice. Here $n_u\in\{0,1\}$. (This model is also equivalent to the Ising anti-ferromagnet in a constant magnetic field.)
Let us consider $0\leq \mu\leq 4U$, where at low temperatures, the repulsion dominates over the chemical potential. Then the ground states are a checkerboard, where we pick a sub-lattice with $n=1$. These two states, corresponding to the two sublattices,  maximize the number of occupied sites, and thus $\mu \sum n$, while avoiding any occupied nearest neighbors.  
The two checkerboard states spontaneously break translation symmetry and are transformed into each other if we shift by one lattice site. This is a solid phase. By contrast, as $T\rightarrow\infty$, the ensemble \eqref{eq:thermalensemble} is uniform: $\mathbb{P}(\mathbf{n}) = 2^{-L^2}$. Here $L$ is the number of vertices along a side of the square.  This ensemble is disordered: all configurations are equally likely.  In fact, this last statement is true at $T=\infty$ ($\beta=0$) \emph{independently of} $H$, since all $\mathbf{n}$ are valid configurations.  This is the heart of a no-go theorem on entropic order as $T\rightarrow \infty$ (see SM). Indeed, many familiar systems observed in nature are disordered at high $T$.

For $\mu$ outside the range $[0,4U]$, the model has no phase transitions and the zero-temperature ground state is unique. 

%If $\mu<0$ and $U\gg|\mu|$, then as $T\rightarrow 0$, the ensemble \eqref{eq:thermalensemble} is dominated by the states with smallest $H$, which turn out to be a pair of checkerboard states: 

\emph{High-temperature order.}--- We are going to explore how the no-go theorem above can be avoided. 
Clearly, this is only possible if the thermal distribution \eqref{eq:thermalensemble} does not look uniform as $\beta \rightarrow 0$.  On the $L\times L$ square lattice, one way to achieve this is if there are an infinite number of microstates: e.g. 
%\fedor{converge in weak sense rather then in a strong sense?} \andy{What does this mean? The ensemble does not exist at $\beta=0$ as it's not normalizable. Maybe we can phrase the idea better, but I want to get across that how the limit $\beta\rightarrow 0$ taken matters.  So a point which might seem like semantic math is actually fundamental to the physics.}: we can take $\beta>0$ arbitrarily small, but never exactly zero.  
%To do this, we will allow 
$n_v \in \lbrace 0,1,2,\ldots \rbrace$ can take any non-negative integer value. Just as there is not a uniform distribution over the integers, a uniform $\beta=0$ ensemble will then not exist.  We choose  \begin{equation}
    H = U \sum_{u\sim v} n_u^2 n_v^2 + \sum_v n_v.\label{eq:numericmodel}
\end{equation}
Note the similarity with~\eqref{IntroModel} where we have fixed $\mu=-1$. 
Our model~\eqref{eq:numericmodel}
is well defined; it has finitely many states $\mathbf{n}$ obeying $H(\mathbf{n})\le E$ for any finite $E$. The repulsion $U$ between particles at adjacent sites can grow slower than quadratically (but faster than linearly) in the $n_u$, without changing the conclusions.  

Intuitively, as $\beta \rightarrow \infty$, the dominant states in the thermal ensemble will have $n_v=0$ on most sites, since each particle costs energy.   For a large but finite $\beta\gg 1$, some finite fraction $\sim \mathrm{e}^{-\beta}$ of sites will be occupied, most with a single particle.  Since this fraction of sites is small, the occupied sites can essentially be drawn randomly, so there is no long-range order.  This is a gas phase. %\Zohar{But why any beta smaller than infinity? i think only for large beta}  %i'll fix...ur right

We now claim that for sufficiently small $\beta<\beta_{\mathrm{c}}$, the model is an entropic ordered solid.  To justify this claim, notice that if one site is occupied but none of its neighbors are, the typical number of particles $\bar n_1$ obeys $\beta \bar n_1 \sim 1$, or $\bar n_1\sim T$.  In contrast, if two adjacent sites both have $\bar n_2$ particles, $\beta \bar n_2^4 \sim 1$ or $\bar n_2 \sim T^{1/4}$.  If we consider the checkerboard arrangement from before, we can occupy half of the sites leading to partition function $\bar n_1^{L^2/2}\sim T^{L^2/2}$, which is much larger than $\bar n_2^{L^2}\sim T^{L^2/4}$ if we consider the disordered state.  This suggests that the dominant contribution to $Z(\beta)$ comes from checkerboard-like states for sufficiently small $\beta$. Therefore the high-temperature phase  is a solid phase that spontaneously breaks the lattice translational symmetry. 
%imagining a chess board, we can pile up particles on either the white or the black squares.  There is no reason to prefer one or the other, but this solid will spontaneously pick one of the two at high temperatures.

\begin{figure}
    \centering
    \includegraphics[width=\linewidth]{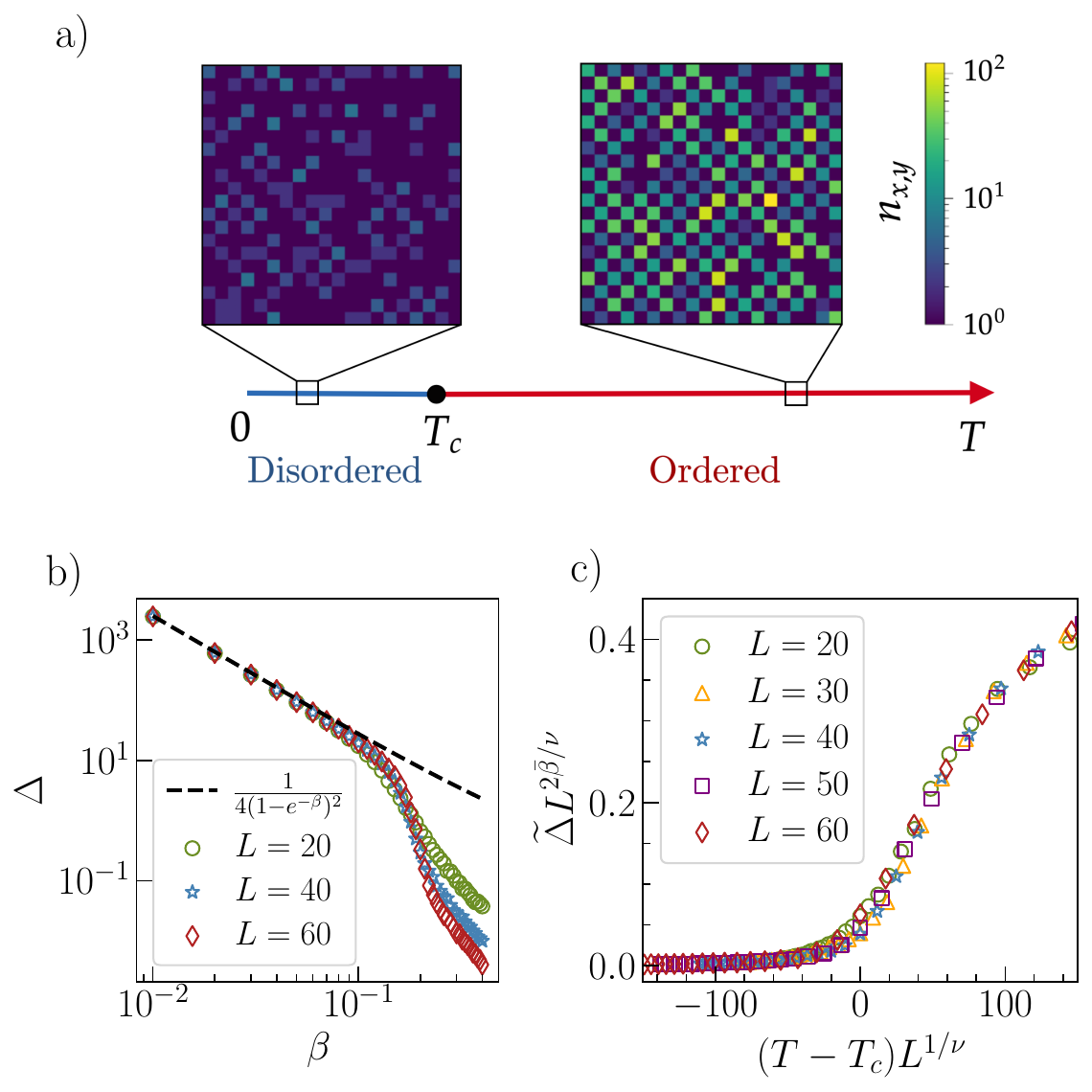}
    \caption{(a) The phase diagram of the classical bosonic lattice gas \eqref{eq:numericmodel}.  We choose two points $T=1/\beta=2$ and $T=20$ and show the corresponding late-time snapshots of the model on a $20\times 20$ square lattice simulated using the Monte Carlo algorithm. The color on each site denotes the number of particles on a site, with brighter color representing more particles.
    %For more visibility, we use a logarithmic colormap, and add 1 to the particle number on each site. 
    (b) The square of the particle density difference of sublattices $A$ and $B$ vs $\beta$ for different system sizes on a log-log scale. The black dashed line corresponds to the theoretical prediction assuming a defect-free solid. (c) A finite-size scaling analysis of the order parameter $\widetilde{\Delta}$, defined analogously to $\Delta$ but with $n_{(x,y)}$ replaced by $1-\delta_{n_{(x,y)},0}$, is consistent with the Ising universality class. For all simulations, we take the interaction strength $U=1$, which gives the critical point $T_{\mathrm{c}}=1/\beta_c\approx1/0.19$.}
    \label{fig:boson}
\end{figure}

An analytic formula for $Z(\beta)$ is not known to us, but we can deduce the phase diagram by numerical Markov Chain simulations \cite{Metropolis1953,Hastings1970}, using a classical Gibbs sampler to (approximately) sample microstates $\mathbf{n}$ with probability \eqref{eq:thermalensemble}.  The results are summarized in Figure \ref{fig:boson}a, where typical states clearly are disordered at low temperature and ordered at high temperature, with the order manifesting in the anticipated checkerboard pattern. In statistical physics, we can more quantitatively diagnose the presence of order by calculating the order parameter
\begin{equation}
     \Delta =\frac{1}{L^4}\left\langle\left( \sum_{x,y=1}^L(-1)^{x+y}n_{(x,y)}\right)^2\right\rangle = \frac{\langle (N_{\mathrm{A}}-N_{\mathrm{B}})^2\rangle}{L^4}.
\end{equation}
which counts the imbalance between the two checkerboards, parameterized by whether $x+y$ is an even or odd integer.  The factor $L^{-4}$ in front ensures that $\Delta>0$ as $L\to\infty$ in a solid phase, while $\Delta=0$ as $L\rightarrow\infty$ in a disordered phase. Fig.~\ref{fig:boson}b demonstrates that $\Delta$ takes an $L$-independent value for $\beta\lesssim \beta_{\mathrm{c}}\approx0.19$ at $U=1$, suggesting that $\beta_{\mathrm{c}}\approx0.19$ is the critical temperature separating order and disorder. We expect that the universality class of this transition matches the two-dimensional Ising model, which we confirm by a standard scaling analysis in Fig. \ref{fig:boson}c (see SM for details).

% The data collapse at the critical point shows that it satisfies the finite-size scaling
% \begin{equation}
%     \widetilde{\Delta}=L^{-2\beta/\nu}f[(T-T_{\mathrm{c}})L^{1/\nu}],
% \end{equation}
% where $\beta=1/8$ \andy{maybe use $\bar\beta$ or something to avoid using the same symbol twice?} and $\nu=1$ are the critical exponents of the $2d$ Ising model. 
% \andy{add a bit more text.  ising universality class.  add figure.}  $\beta=1/8$ (not to be confused with the inverse temperature) 
%We conclude that this model has an entropic ordered solid at high $T$, and the phase transition from disorder to order belongs to the Ising universality class. As we explained, physically this is because  by ordering into a checkerboard, we enable fluctuations of $\bar n_1 \gg \bar n_2$ bosons per site, which allows to increase the entropy significantly in the ordered state. 

An interesting limit that we can say more about is to take $U\to\infty$ with fixed $\beta$. 
This is a variant of the hardcore lattice gas, which is typically defined as follows: letting $n_u\in\lbrace0,1\rbrace$ \cite{baxter_1980_hardsquare,baxter1982},
\begin{equation}\label{hardcore} 
Z = \sum_{\{n_u\}\text{ obeying constraint}}  z^{\sum_u n_u} = \sum_{M=0}^{N^2/2} F_{M}z^M
\end{equation}
where $z=\mathrm{e}^{\beta \mu}$, and valid configurations in the partition function obey $n_un_v=0$ for any $u\sim v$ (no neighboring sites are both occupied).  Notice that $\frac{N^2}{2}$ is the maximal number of total particles.
%\Zohar{Furthermore the number of particles is $n_u\in\{0,1\}$ in the standard version of the hardcore gas. Thus, $N^2/2$ is the maximal total number of particles given the constraints.} 
We have defined $F_{M}$ as the number of ways to occupy $M$ sites on the $L\times L$ lattice obeying the constraint. The model~\eqref{hardcore} is the  $U\to\infty$ limit of the model~\eqref{IntroModel}. 
%and the partition function is as usual $Z=\sum_{\{n\}} e^{\beta\mu\sum_u n_u}$.

If $\mu>0$ then the low temperature phase ($z\rightarrow\infty$) is a solid and the high temperature phase $(z\rightarrow 1)$ is a gas. If $\mu<0$ there is a gas at all temperatures. The phase transition at $\mu>0$ on the square lattice is in the Ising universality class
and occurs at $z\approx3.79$ \cite{baxter_1980_hardsquare,Fernandes_2007}. 
%From our perspective, the high-temperature limit of this model is clearly $z=1$, which is disordered.  Sometimes, however, authors interpret $Z$ as already an infinite-temperature partition function in a finite-density ensemble, where the choice of $z$ fixes the average density.  In this setting, infinite-temperature order is clearly possible, but it relies fundamentally on the exact hardcore constraint, along with restricting to a certain symmetry/charge sector.

% A simpler (yet somewhat artificial) model that behaves similarly is a variant of the hardcore lattice gas model.  The classic setup is that of particles on an $L\times L$ square lattice, where if a particle is present at some vertex then no additional particles can be present at the same location or adjacent vertices to it. 
%The Hamiltonian and the partition function are simply 
%\begin{equation}\label{hardcore} 
%H=-\sum_u\mu n_u, \quad Z = \sum_{\{n_u\}\text{ obeying constraint}}  z^{\sum_u n_u} 
%\end{equation}
%where $z=\mathrm{e}^{\beta \mu}$.
%and the partition function is as usual $Z=\sum_{\{n\}} e^{\beta\mu\sum_u n_u}$.
%If $\mu>0$ then the low temperature phase ($z\rightarrow\infty$) is a solid and the high temperature phase $(z\rightarrow 1)$ is a gas. If $\mu<0$ there is a gas at all temperatures. The phase transition occurs at $z\approx3.79$ and is in the Ising universality class \cite{}.

Now we return to the  $U\rightarrow\infty$ limit of our model~\eqref{eq:numericmodel}. Because we are no longer restricted to having 0 or 1 particles per site, we now find \begin{equation}
Z(\beta) = \sum_{M=0}^{N^2/2} F_M \left(\mathrm{e}^{-\beta}+ \mathrm{e}^{-2\beta}+\cdots\right)^M = \sum_{M=0}^{N^2/2} \frac{F_M}{(\mathrm{e}^\beta-1)^M}.    
\end{equation}
% An interesting variant of this model is to allow an arbitrary number of particles on each site, but as soon as there are any particles on a given site, no adjacent sites can have particles.
% Denote $F(M,N)$ the number of ways to choose $M$ occupied sites out of $N=L^2$ total sites. The partition function of~\eqref{hardcore} is $Z=\sum_MF(M,N)z^M$, with $z=e^{\mu\beta}$. In our model with unrestricted number of particles per site the result is similar: 
% \begin{equation} Z=\sum_{M}F(M,N)(z+z^2+\cdots)^M=\sum_{M}F(M,N)\left({z\over 1-z}\right)^M~. \end{equation}
% We can thus read out the phases of this model using the traditional hardcore lattice gas with the map $z\rightarrow \frac{z}{1-z}$.
%$$z\to {z\over 1-z}~.$$
% We have to take now $\mu<0$ or $z<1$ since the Hamiltonian is otherwise not well defined.
% For very low temperature $z\to0$ we see that ${z\over 1-z}\to 0$ and hence the model is in the fluid phase, as expected, with a few particles in the system. For high temperatures $z\to 1^-$ we have ${z\over 1-z}\to +\infty$ and hence we are in the solid phase! This is physically reasonable, as there are many more ordered states in the system than disordered ones, similarly to our argument in~\eqref{eq:numericmodel}.
We see that the phase diagram is exactly flipped relative to the standard hardcore lattice gas:  now $z=(\mathrm{e}^{\beta}-1)^{-1}$ is small at low temperature ($\beta \rightarrow \infty$), and large at high temperature ($\beta\rightarrow 0$). Therefore the model describes a low-temperature gas and a high-temperature entropic-ordered solid. 
%This  model realizes an entropic-ordered solid. at high temperature.  

The mechanism underlying the formation of this high-temperature solid is that sacrificing the mobility of the particles and placing them in a checkerboard pattern  allows to gain  entropy due to the fluctuations on site.
In fact, it is not necessary to allow an arbitrary number of particles at each site for this mechanism to work! If we denote the maximum allowed number of particles at each site by $K$, so long as $K\ge 4$, we still find a solid at high temperature on the rectangular lattice.  

This model with finite $K$ (and $U=\infty$) has finitely many states, but still leads to high temperature order, because the configuration space is not a direct product, due to the hardcore constraint.
Similarly, one can construct other examples of high-temperature order in systems at a fixed charge sector. For example, the model~\eqref{hardcore} at a fixed density will have high-temperature order for sufficiently large density (that corresponds to fixed $\beta\mu$ with $\beta\to0$).  We focus on realizing entropic order without such constraints.

%Of course, as we have seen, this conclusion relies on the exact constraint when $U=\infty$.
%Therefore, even with finitely many degrees of freedom per site, entropic order is possible. 
% The partition function is (assuming even $L$) \begin{equation}
 %   Z(\beta) = \sum_{N=0}^{L^2/2} C_N \left(\frac{1}{\mathrm{e}^{\beta}-1}\right)^N 
%\end{equation}
%where $N$ counts the number of occupied sites and $C_N$ counts the number of configurations of occupied sites obeying the hardcore constraint.  It is known that this model exhibits a phase transition from a liquid to a solid on the square lattice when $1/(\mathrm{e}^{\beta_{\mathrm{c}}}-1) = z_{\mathrm{c}} \sim $ \andy{yiqiu?}, with the solid phase arising for $z>z_{\mathrm{c}}$.   The crucial thing is that as $\beta \rightarrow 0$, the fugacity $z\rightarrow \infty$, in contrast to what would have happened if we had restricted the particle occupation number to be 0 or 1 (as was usually assumed in the literature).  
%Formally, the hardcore model is given by the $U\rightarrow \infty$ limit of~\eqref{eq:numericmodel}.

\emph{Exactly solvable models.---} To illustrate that \eqref{eq:numericmodel} is just one of many simple models with entropic order, we now describe some exactly solvable models, starting with a classical model of ferromagnetism at high $T$, based on the Ising model.  Let $s_v \in \lbrace \pm 1\rbrace$ be spins sitting on the vertices of a square lattice, 
%while $n_{uv}\in\lbrace 0,1,2,\ldots\rbrace$ are ``bosons" sitting on the edges; take \begin{equation}
 %   H = \sum_{u\sim v}\left(a-bs_us_v\right)n_{uv} \label{eq:Isingboson}
%\end{equation}
%where $a>b>0$.  We now evaluate the partition function by exactly summing over $n_{uv}$, and find \begin{equation} 
 %   Z(\beta)= \sum_{s_v}\sum_{n_{uv}}\mathrm{e}^{-\beta H} = \sum_{s_u=\pm 1} \mathrm{e}^{-A+Bs_u s_v} \label{eq:ZIsing}
%\end{equation} where we define \begin{subequations} \label{eq:ZIsing2}
%    \begin{align} 
        %\mathrm{e}^{-2A} &= \frac{1}{\left(1-\mathrm{e}^{-\beta(a+b)}\right)\left(1-\mathrm{e}^{-\beta (a-b)}\right)}, \\
        %\mathrm{e}^{2B} &= \frac{1-\mathrm{e}^{-\beta(a+b)}}{1-\mathrm{e}^{-\beta (a-b)}}. \label{eq:Beq}
        %\end{align}
%\end{subequations}
%Equation \eqref{eq:ZIsing} takes the form of an Ising partition function at effective inverse temperature $B(\beta)$, where $B(\beta)$ is a \emph{decreasing} function of temperature with $B(0)=\frac{1}{2}\log \frac{a+b}{a-b}$ and $B(\infty)=0$.  Using the exact critical temperature of the Ising model on the square lattice, we conclude that if $\frac{a}{b}>1+\sqrt{2}$, the system \eqref{eq:Isingboson} has entropically ordered magnetism (long-range order in $s$).   Again, this model has entropic order: when the discrete spins form a ferromagnet, they enable vastly more fluctuations in the number of bosons $n_{uv}$.  The %reduction in %entropy from the %spins is outweighed %by the gain in %entropy from %$n_{uv}$ %fluctuations.
while $n_{uv}\in\lbrace0, 1,2,\ldots\rbrace$ are ``bosons" sitting on the edges; take \begin{equation}
    H = \sum_{u\sim v}\left(a-bs_us_v\right)(n_{uv}+1) \label{eq:Isingboson}
\end{equation}
where $a>b>0$. This model has a twofold degenerate ground state with $s_u=1$ or $s_u=-1$. The model therefore has the usual magnetic order at low temperatures.  

We now evaluate the partition function by exactly summing over $n_{uv}$, and find \begin{equation} 
    Z(\beta)= \sum_{s_v}\sum_{n_{uv}}\mathrm{e}^{-\beta H} = \sum_{s_u=\pm 1} \mathrm{e}^{-A+Bs_u s_v} \label{eq:ZIsing}
\end{equation} where we define \begin{subequations} \label{eq:ZIsing2} 
    \begin{align} 
        \mathrm{e}^{-2A} &= \frac{\mathrm{e}^{-2\beta a}}{\left(1-\mathrm{e}^{-\beta(a+b)}\right)\left(1-\mathrm{e}^{-\beta (a-b)}\right)}, \\
        \mathrm{e}^{2B} &=\mathrm{e}^{2\beta b} \frac{1-\mathrm{e}^{-\beta(a+b)}}{1-\mathrm{e}^{-\beta (a-b)}}. \label{eq:Beq}
        \end{align}
\end{subequations}
Eq. \eqref{eq:ZIsing} takes the form of an Ising partition function at effective inverse temperature $B(\beta)$. At $\beta\to\infty$, $B\to \infty$ and the model is in the ordered phase, as anticipated.
At $\beta\to 0^+$ we have that 
%where $B(\beta)$ is a \emph{decreasing} function of temperature with 
$B(0)=\frac{1}{2}\log \frac{a+b}{a-b}$.
Using the exact critical temperature of the Ising model on the square lattice, we conclude that if $\frac{a}{b}>1+\sqrt{2}$, the system \eqref{eq:Isingboson} has ferromagnetism (long-range order in $s$) at all temperatures. Indeed one can check that $B$ is a monotonically decreasing function of temperature so one never leaves the magnetically ordered phase. The high temperature magnetism arises from entropic order; 
%Again, this model has entropic order as $T\rightarrow \infty$: 
when the discrete spins form a ferromagnet, they enable vastly more fluctuations in the number of bosons $n_{uv}$.  The reduction in entropy from the spins is outweighed by the gain in entropy from $n_{uv}$ fluctuations. If one is only interested in obtaining very high temperature ordered phases,  %transitions, such as magnetism up to very high temperatures, 
it is not necessary to have infinitely many bosons $n_{uv}\in\lbrace 0,1,2,\ldots\rbrace$: limiting $n_{uv} \le K$ is sufficient. The same comments apply for the constructions below.

 It is possible to use similar  ``link bosons" $n_{uv}$ to construct ordered high temperature phases in a broad variety of other models.  For example, adding link bosons to a three-dimensional classical XY model, we can obtain  superfluidity at arbitrarily high temperatures.
Similarly, adding bosons to the four-dimensional toric code \cite{alicki2008thermal}, we can construct examples of entropic topologically ordered states at high temperature. 
This construction demonstrates a clear loophole in the recent theorem that all quantum lattice models have no quantum entanglement at sufficiently high $T$ \cite{Bakshi:2024cqr}.  As in the classical setting, this theorem assumed that the local Hilbert space was finite-dimensional, an assumption violated by the bosons above. Details can be found in the SM. 
%Although in bosonic Hubbard-type models it is possible to prove theorems analogous to those for quantum spin systems \cite{} (there is no high-$T$ order), our example clearly illustrates that such theorems cannot hold in generality: quantum statistical mechanics can be qualitatively different with bosons. 

%further generalize our constructions to quantum many-body models.  In quantum mechanics, %Hamiltonian $H$ becomes an operator on a Hilbert space (see SM), and 
%the analogue of \eqref{eq:thermalensemble} is the Gibbs state $\rho = \mathrm{e}^{-\beta H}/Z(\beta)$, with $Z(\beta) = \mathrm{tr}\left(\mathrm{e}^{-\beta H}\right)$.  We can readily construct uniquely quantum entropic ordered phases, with the most striking example having \emph{topological order} \cite{finiteTtopological}, where the thermal state cannot be prepared in a low-depth circuit from an unentangled state.  An example of topological order at large but finite $\beta$ is the four-dimensional toric code \cite{}.  Quantum states with this level of computational complexity are useful for quantum error correction, since heuristically, each logical qubit cannot be destroyed by a small amount of environmental noise of any kind.  Many such error-correcting codes can be described by stabilizer Hamiltonians $H = -\sum_a S_a$, where $S_a$ form a collection of commuting operators ($[S_a,S_b]=0$) -- see SM for details.  We can add ``link bosons" for each $S_a$ to build a model of entropic topological order as $T\rightarrow \infty$. 

\emph{Quantum field theory.---} 
Models of entropic order also exist in continuous space, i.e. in quantum field theories (QFTs)  as first considered in \cite{weinberg1974}. To be able to push entropic order all the way to infinite temperature in QFT, the QFT has to be UV-complete, i.e. exist independently of an underlying lattice model. 
%\Zohar{We have reviewed recent progress in the introduction and we will go into some more detail below.} 
%Some models in fractional dimensions were found in \cite{Chai:2020onq,Chai:2020zgq,Liendo:2022bmv,Chai:2020hnu,Chaudhuri:2020xxb,Bajc:2020gpa,Nakayama:2021fgy,Chaudhuri:2021dsq,Agrawal:2021alq} and more recently a local and unitary model in $d=2$ was found in \cite{Hawashin:2024dpp,Komargodski:2024zmt}. 
Our goal here is to show that the QFTs of \cite{Chai:2020onq,Chai:2020zgq,Komargodski:2024zmt} are entropically ordered at high temperature: ordered states carry more entropy than the disordered ones. 

For simplicity we will focus on QFTs in $2<d<3$ spatial dimensions (see SM for details).
A similar, albeit more complicated, analysis can be carried out in $d=2$ as well (the conclusions remain the same). 

%mechanism in this model is that of entropic order, i.e. that ordered states have a higher entropy than disordered ones.

%It is natural to use Feynman's path integral to express the partition function as \begin{equation}
 %   Z(\beta) = \int \mathrm{D}\phi \mathrm{e}^{-\int\limits_0^\beta \mathrm{d}\tau \int \mathrm{d}^dx \mathcal{L}(\phi,\partial \phi,\ldots)},
%\end{equation}
%where $\mathcal{L}$ is the Euclidean time Lagrangian density and, here, $\phi$ represents all quantum fields in the theory.  Notice that we compactify imaginary time $\tau \sim \tau+\beta$ into a circle, whose circumference is the inverse temperature.  Conceptually, an elegant way to evaluate this path integral is to imagine separating $\phi = \bar \phi + \tilde \phi$ into a space-time independent constant $\bar \phi$, and fluctuations $\tilde \phi$, and to approximate that we can evaluate $Z(\beta)$ by expanding $\mathcal{L}$ to quadratic order only in $\tilde \phi$.  Upon doing so, we obtain \begin{equation}
 %   Z(\beta) \approx \int \mathrm{d}\bar\phi \mathrm{e}^{-\beta V \mathcal{F}(\bar \phi)}
%\end{equation}
%where $\mathcal{F}(\bar \phi)$ is the (1-loop effective) free energy density arising from quantum and thermal fluctuations around the ``vacuum" $\bar \phi$.  While $\mathcal{F}(\bar \phi)$ does not always exist, it is well-defined near the stable minima of $\mathcal{F}$ which dominate $Z(\beta)$ \cite{}.

The model that we study contains a boson $\psi$ and a vector of $N$ bosons $\vec \phi$ with Lagrangian (in Euclidean signature) %Adopting known results \cite{}, we show that given  
\begin{equation} \label{eq:QFTL}
    \mathcal{L} = \frac{1}{2}(\partial \psi)^2 + \frac{1}{2}(\partial \vec{\phi})^2 + \frac{\lambda}{4N} \left(\vec{\phi}^2 - \psi^2\right)^2,
\end{equation}
%where $\phi^a$ $(a=1,\ldots,N)$ is an $N$-component field and $\psi$ will exhibit entropic order, 
This model has local interactions and the energy is bounded from below. 
The classical Hamiltonian has a degeneracy of ground states with $\vec{\phi}^2=\psi^2$ but this degeneracy is lifted quantum mechanically \cite{Chai:2020onq,Chai:2020zgq,Komargodski:2024zmt}, and there is a unique ground state at zero temperature in the full theory. In fact the model~\eqref{eq:QFTL} leads to an interacting conformal theory at zero temperature. That conformal theory is multi-critical since several relevant parameters are tuned to zero. We investigate the high temperature behavior of the multi-critical fixed point, since that fixes the high temperature behavior of the nearby ordered and disordered zero temperature phases as well.

%(Strictly speaking, in 2+1 dimensions one also has to consider the marginal $\psi^6$ coupling, but it does not affect our discussion so we ignore it.)

Our goal is to calculate the 
the free energy of a configuration with average field configurations  $\bar \phi^a=(\bar\phi,0,\ldots,0)$ and $\bar \psi$. Using standard techniques of thermal field theory, in the large $N$ limit, the answer is given by
\begin{gather}
    \frac{\mathcal{F}}{N} \approx -c_1 T^{d+1}+c_2T^{-\frac{2}{d-2}}\left(c_3T^{d-1}+\phi^2-\psi^2\right)^{\frac{d}{d-2}} + \cdots \label{eq:mainfreeenergydensity}
\end{gather}
where $c_{1,2,3}>0$ are constants and $\cdots$ stand for terms which are sub-leading at large $N$.
$\mathcal{F}$ above still has a manifold of thermal minima $c_3T^{d-1}+\phi^2-\psi^2=0$. This degeneracy is resolved once one includes further corrections in the $1/N$ expansion. The final answer is that \cite{Chai:2020onq,Chai:2020zgq,Komargodski:2024zmt} $\mathcal{F}$ is minimized when  $\bar\phi^2=0$ and $\bar\psi^2= c_3 T^{d-1}$.  Since $\bar\psi^2>0$, the $\mathbb{Z}_2$ symmetry $\psi \rightarrow -\psi$ of \eqref{eq:QFTL} is spontaneously broken at any $T>0$.
%The true minimal value of $N$ at which entropic order occurs could be larger than $N=3$ []. 
To see that this ordered state maximizes the entropy density we calculate it explicitly: at $\bar\phi=0$, %ANDY: this is because the formula is SO LONG otherwise...
% \begin{align}
%     \cS &= - \left(\frac{\p \cF}{\p T}\right)_{\bar\phi,\bar\psi}=(d+1)(N+1)c_1 T^d\nonumber\\
%     &  -(d-1)c_2 \frac{\lambda}{N} T^{d-2}\left((N+1)\bar\phi^2 - (N-3)\bar\psi^2\right).
% \end{align}
\begin{align}
    \mathcal{S} &= - \left(\frac{\p \cF}{\p T}\right)_{\bar\phi,\bar\psi} = (d+1)c_1NT^d \\
    &- \frac{T^{-\frac{d}{d-2}}c_2(c_3T^{d-1}-\bar\psi^2)^{\frac{2}{d-2}}}{d-2}\left(2 \bar\psi^2+ (d^2 - d - 2)c_3T^{d-1}\right). \notag
\end{align}
Indeed, ordering $\bar\psi$ \emph{increases} the overall entropy.

%It is important to note that that the above estimate $N>3$ is not reliable since we have performed a large $N$ expansion to leading and sub-leading order. All we can say that the above holds for some finite, large enough, $N$ but more work is required to figure out when entropic order kicks in. 

%\andy{this does not introduce the $\sigma$ field, is it possible to ``integrate it back out"?  I think that calculation would be simpler to explain as I tried above}   Formally this model is not UV-complete as a quantum field theory.  However, using the UV-complete conformal field theories obtained in \cite{}, we are able to similarly evaluate $\mathcal{F}$ and observe entropic order in the symmetry-broken phase with $\langle \psi^2\rangle >0$ at high temperature (see SM).  The mechanism underlying entropic order in this QFT is analogous to the solvable model \eqref{eq:Isingboson}.

\emph{Superconductivity.---} Now that we understand how to build quantum field theories and lattice models with entropic order, we discuss a model of high-temperature superconductivity.  The standard Bardeen-Cooper-Schrieffer (BCS) model for superconductivity \cite{bcs} contains a finite density Fermi surface of spin-$\frac{1}{2}$ electrons $\psi_{\uparrow,\downarrow}$ with an attractive two-body interaction, leading to a spontaneously broken U(1) symmetry, measured by a non-vanishing order parameter $\Delta = \langle \psi_\uparrow \psi_\downarrow\rangle$.  At low temperatures relative to a cutoff energy $\omega_*$, which can be large compared to room temperature, we find superconductivity whenever the BCS gap equation (see SM) has a solution:
%Consider spinful fermions $\psi_{\uparrow,\downarrow}$ at a finite chemical potential $\mu$ with an attractive interaction $\propto g$.
% onset of the superconductivity is due to the instability of the Fermi liquid toward condensing the Cooper pair field $\abs{\Delta}^2 = g \abs{\langle \psi_{\downarrow} \psi_{\uparrow} \rangle}^2  \neq 0$, which happens when, according to the BCS theory, $T<T_{\mathrm{c}}$ with $T_{c}\sim\omega_* \exp(-1/g\nu)$ and $\nu$ is the density of state near the Fermi energy and $\omega_*$ is the Debye frequency. Let us generalize this BCS theory by coupling to a large $N$ scalar field theory just like  in the previous section. (In practice, $N=3$ scalars could be enough, as above.)
\begin{align}\label{eq:gap}
    \frac{1}{g_{\mathrm{eff}}} = \nu \int\limits_0^{\beta \omega_*} \ud x ~ \frac{\tanh\left(\frac{1}{2}\sqrt{x^2 + (\beta\Delta)^2}\right)}{\sqrt{x^2 + (\beta\Delta)^2}},
\end{align}
where $\nu$ is the fermionic density of states, and $g_{\mathrm{eff}}$ is the effective interaction strength. 
Regardless of the value of $\Delta$, the integral above goes to zero as $\beta\to 0$. Therefore, at sufficiently high temperature, one finds no solution if $g_{\mathrm{eff}}$ is temperature independent, and thus there is no superconductivity at high temperature.   Crucially, we can build a model where $g_{\mathrm{eff}}(T)$ \emph{increases} with temperature.   Keeping details in the SM, we write a theory of $N\gg 1$ critical bosons coupled to $N\gg 1$ spin-$\frac{1}{2}$ fermions,  similar to \eqref{eq:QFTL}, but where $\psi^2$ is replaced by $|\Delta|^2$.  We find entropic order, which manifests in a $g_{\mathrm{eff}}$ which grows at higher temperature.  This temperature dependence of $g_{\mathrm{eff}}$ enables superconductivity for all $\beta\omega_*\gtrsim 1$ (the effective field theory does not make sense at higher temperatures). This is in contrast to the usual case where superconductivity persists up to $\beta_{\mathrm{c}}\sim \omega^{-1}_* \mathrm{e}^{1/g_{\mathrm{eff}}\nu}  \gg \omega^{-1}_*$, i.e. we have found an exponential increase in $T_{\mathrm{c}}$.

%These conclusions persist when we take into account fluctuations of a U(1) gauge field coupled to the charged fermions.
%$g_{\mathrm{eff}}$ is the effective interaction strength between fermions.  Ordinarily, $g_{\mathrm{eff}}$ is a constant, and 
%where $\sigma$ is the Hubbard–Stratonovich field that captures the thermal mass of the $N$ scalars, and $t>0$ is the coupling constant. If $t=0$, \eqref{eq:gap} becomes the conventional BCS gap equation, which leads to $T_{\mathrm{c}}$. However, the correct solution of the $N$ scalars model has  $\sigma\neq 0$ and increases with $T$. Then there exists a solution $\abs{\Delta}^2\sim t^{-1}T^{d-1}>0$ for any high $T$ (SM). %Recall that $\sigma$ is the thermal mass for the large-$N$ scalar, 
%The shift $1/g\to 1/g - \sigma$ can  be regarded as inducing an effective $T$-dependent attractive interaction.     

%material parameters needed for such a mechanism, which may well be challenging to realize experimentally.  A practical challenge may be the fine-tuning of parameters to reach a critical point.

\emph{Conclusion.---} We have described entropic order, whereby typical high-energy states, of either classical or quantum systems, can exhibit long-range order, and/or quantum entanglement. This counterintuitive idea is possible because sometimes ordering a subset of degrees of freedom enables many more possible microstates for the rest.

We have demonstrated this concept for lattice gases which turn into solids at high $T$, magnets that remain magnetic at high $T$, persistent superfluidity, topological order, and high-$T_{\mathrm{c}}$ superconductivity.  Entropic order was also seen to explain the recent demonstration of $T\rightarrow \infty$ order in QFT.

An important ingredient in our construction of high-temperature superconductivity were interacting bosons, which, under the circumstances we described, lead to entropically-driven superconductivity. This is in contrast to simply enhancing the effective zero-temperature coupling~\cite{kivelson}. It would be very interesting if such ideas are realizable.

\section*{Acknowledgements}
We thank Sarang Gopalakrishnan, David Huse, and Leo Radzihovsky for interesting discussions.  This work was supported by the National Science Foundation under CAREER Grant DMR-2145544 (XH, AL) and by the Air Force Office of Scientific Research under Grant FA9550-24-1-0120 (YH, AL).

\onecolumngrid

\newpage

\appendix

\renewcommand{\thesubsection}{\Alph{section}.\arabic{subsection}}

\section{Classical statistical physics}\label{app:classical}
In this appendix we present a somewhat more mathematical discussion of classical statistical physics, the existing no-go theorems  on high-temperature order in lattice models, and how to avoid them.

\subsection{Known definitions and results}

In this paper, we typically consider the \emph{configuration space} $\Omega$ to be a countable set.  In an \emph{unconstrained many-body model} with $N$ degrees of freedom, we have \begin{equation}
    \Omega = \Omega_1 \times \cdots \times \Omega_N. \label{eq:unconstrainedspace}
\end{equation}
For simplicity we always take $N$ to be finite; the limit $N\rightarrow\infty$ at the end of a calculation is the \emph{thermodynamic limit}.   An intuitive notion of (Hamming) distance $\mathsf{d}:\Omega\times\Omega \rightarrow \mathbb{Z}^+$ on the configuration space is that $\mathsf{d}(\omega_1,\omega_2)$ counts the number of entries in which the $N$-component tuples $\omega_{1,2}$ differ.  Given two subsets $A_{1,2}\subset \Omega$, we can define $\mathsf{d}(A_1,A_2) = \min_{\omega_1\in A_1, \omega_2\in A_2} \mathsf{d}(\omega_1,\omega_2)$. 

The Hamiltonian (energy) is a function $H:\Omega \rightarrow \mathbb{R}$.  The probability of observing microstate $\omega \in \Omega$ is given by \begin{equation}
    \mathbb{P}_\beta(\omega) = \frac{\mathrm{e}^{-\beta H(\omega)}}{Z(\beta)} \label{eq:appendixZ}
\end{equation}where $\beta=1/T$ and \begin{equation}
    Z(\beta) = \sum_{\omega \in \Omega} \mathrm{e}^{-\beta H(\omega)} =: \mathrm{e}^{-\beta F(\beta)}.
\end{equation}
We assume that $H$ is chosen such that $Z(\beta)$ is finite for all $0<\beta<\infty$, even if $\Omega$ is countably infinite.  Examples of this case will be discussed below.  A system has \emph{symmetry} group $G$ if there is a group action of $G$ on $\Omega$ such that for all $g\in G$ and $\omega\in\Omega$, $H(g\cdot \omega)=H(\omega)$.\footnote{Note that there exist zero-temperature symmetries that cannot be preserved by the thermal ensemble, such as boost symmetry \cite{Alberte:2020eil,Komargodski:2021zzy}.}

We will further assume that \begin{equation}
    H(\omega) = \sum_{S\subset \lbrace 1,\ldots, N\rbrace : |S|\le k} H_S(\omega_S) \label{eq:assume1}
\end{equation}
for some O(1) number $k$; here $\omega_S$ corresponds to the subset of the tuple $\omega$ associated with sites in $S$. In computer science, these systems are called $k$-local.  For simplicity, we focus on systems which obey $\max_S |H_S| = \mathrm{O}(1)$, and
\begin{equation}
    \max_{i\in \lbrace 1,\ldots ,N\rbrace} |\lbrace S : i\in S\rbrace |  = \mathrm{O}(1). \label{eq:assume2}
\end{equation}

The most ``liberal" definition of an \emph{ordered phase of matter} is a Hamiltonian $H$ on a configuration space $\Omega$, along with a temperature $1/\beta$ at which we can construct a collection of disjoint sets $A_1,\ldots, A_R \subset \Omega$ such that $\mathsf{d}(A_i,A_j)$ grows with $N$ faster than any constant,\footnote{This is usually denoted as $\omega(1)$, not to be confused with a microstate $\omega$!} 
i.e. $A_{i}$ and $A_j$ cannot be reached by flipping only a finite number of spins in the thermodynamic limit, and in the thermodynamic limit we observe the system to be in the  $A_1,\ldots,A_R$ with probability 1: \begin{equation}
    \sum_{i=1}^R \mathbb{P}_\beta (A_i) = 1 - \mathrm{o}(1). \label{eq:condensation}
\end{equation} 
This is said to be a \emph{condensation of the Gibbs measure} onto disjoint sets.  Usually, the ``$\mathrm{o}(1)$" in the above formula decays sufficiently rapidly with $N$ (e.g. as a stretched exponential) such that we consider ergodicity to be broken.  More concretely, if we study the dynamics of an arbitrary Markov chain \cite{levin_markovchains} on $\Omega$, with few-body transition rules, we can prove strong bounds on the average time it takes to move from $A_i$ to any other $A_j$.   

In the simplest examples, this measure condensation can be associated with \emph{spontaneous symmetry breaking} (SSB).  For example, let us illustrate the mathematical formalism above in an example well-known to physicists: the two-dimensional Ising model on an $L\times L$ square lattice.  Here we have $\Omega = \lbrace \pm 1\rbrace^{L^2}$: each microstate $(s_{11},\ldots, s_{LL})$ is a ``bit string" of $\pm 1$ spins on every lattice site.  The Hamiltonian is \begin{equation}
    H(\mathbf{s}) := -\sum_{u\sim v}s_u s_v.
\end{equation}
For sufficiently large
\begin{equation}
    \beta > \beta_{\mathrm{c}} = \frac{\log(1+\sqrt{2})}{2},
\end{equation}
one finds that $\mathbb{P}_\beta$ condenses onto two sets $A_+$ and $A_-$, defined as sets on which the majority of spins are $+1$ or $-1$ respectively, \emph{and} in which there is no connected domain wall of length $\ge L$.  One can prove that  \cite{Thomas1989} \begin{equation}
    \mathbb{P}_\beta(A_+) = \mathbb{P}_\beta(A_-) = \frac{1}{2}- c\mathrm{e}^{-bL}
\end{equation}
for some $\beta$-dependent $b,c>0$. 
The two-dimensional Ising model has a $\mathbb{Z}_2$-global symmetry corresponding to $s_v \rightarrow -s_v$ for every vertex $v$, which exchanges $A_+$ with $A_-$.  Therefore we say that there is SSB.

In physics, it is usually far more useful to probe SSB by detecting \emph{long-range order} in two-point correlation functions which -- unlike $\mathbb{P}_\beta$ -- are measurable in experiments.  Using $\langle \cdots \rangle $ to denote expectation values with $\mathbb{P}_\beta$, one finds that \begin{equation}
  \lim_{\mathsf{d}(u,v)\rightarrow\infty } \left[ \langle s_u s_v\rangle - \langle s_u\rangle\langle s_v\rangle\right]  > 0.
\end{equation}
Here we consider the averages in the Gibbs ensemble in the large volume limit without any boundary conditions. For instance, in the Ising model, $\langle s\rangle =0$ always because $s$ is charged under the $\mathbb{Z}_2$ symmetry.  Thus  $\langle s_us_v\rangle$ decays exponentially in the disordered phase and approaches a constant in the ordered phase. If we took the large volume limit with, e.g., fixed spin boundary conditions, then $\langle s_us_v\rangle-\langle s_u\rangle \langle s_v\rangle$ would exponentially vanish in both phases.

With this mathematical and physical background in mind, we can now state a powerful theorem: for any many-body system obeying the assumptions \eqref{eq:assume1} and \eqref{eq:assume2}, so long as $\Omega$ is a finite set which is unconstrained and takes the form of \eqref{eq:unconstrainedspace}, there exists an O(1) number $\beta_*$ such that if $\beta<\beta_*$, there is no long-range order: \begin{equation}
        \langle A_u B_v\rangle - \langle A_u\rangle \langle B_v\rangle \le c \mathrm{e}^{-b \mathsf{d}(u,v)}
    \end{equation}
    for some $c,b>0$.\footnote{This is a special case of quantum results we discuss in Appendix \ref{app:qm}.} 

We do not present the proof of this known result, but conceptually it follows from the cluster expansion, along with the fact that the limit $\beta \rightarrow 0$ of $\mathbb{P}_\beta$ exists.

\subsection{A generic mechanism for classical entropic order}\label{app:classicalexamples}
As we have emphasized in the main text, our main interest in this paper is on systems with entropic order at arbitrarily high temperature.  Clearly, we must break the assumptions of the above theorem, and the most interesting way is to relax the assumption that $\Omega_i$ (and thus $\Omega$) is a finite set.   We have already illustrated in the main text how to do this for a two-dimensional Ising model on a square lattice, whose configuration space is $\lbrace \pm 1\rbrace^{L^2}$.

More generally, consider a finite and unconstrained configuration space $\Omega$.  Suppose that there is some Hamiltonian of the form \eqref{eq:assume1} for which at temperature $\beta=1$, \begin{equation}
    \bar H = \sum_S \bar H_S
\end{equation}
is in a non-trivial phase of matter.  We can easily write down a model where this same phase of matter is realized by entropic order as $\beta \rightarrow 0$.  To do so, for each set $S$, append a ``transmutation boson" $n_S \in \mathbb{Z}^+ = \lbrace 1,2\ldots\rbrace$, and define \begin{equation}
    H = \sum_S \mathrm{e}^{\bar H_S}n_S. \label{eq:A13}
\end{equation}  
The new configuration space is \begin{equation}
    \Omega_{\mathrm{new}} = \Omega \times \bigtimes_S \mathbb{Z}^+. \label{eq:Omeganew}
\end{equation}
We can now evaluate the partition function in an expansion at small $\beta$

\begin{align}
    Z(\beta) = \sum_{\omega \in \Omega} \sum_{n_S=1}^\infty \mathrm{e}^{-\beta H} = \sum_{\omega \in \Omega}\prod_S \frac{\mathrm{e}^{-\beta \exp(\bar H_s)}}{1-\mathrm{e}^{-\beta \exp(\bar H_S)}}= \sum_{\omega \in\Omega}\exp\left(-\sum_S \left[\log \beta + \bar H_S + \frac{\beta}{2}\mathrm{e}^{\bar H_S} + \cdots \right] \right).
\end{align}

%\begin{align}
 %   Z(\beta) = \sum_{\omega \in \Omega} \sum_{n_S=1}^\infty \mathrm{e}^{-\beta H} = \sum_{\omega \in \Omega}\prod_S \frac{e^{-\beta \exp(\bar H_s)}}{1-\mathrm{e}^{-\beta \exp(\bar H_S)}}&=\sum_{\omega \in\Omega}\exp\left(-\sum_S \left[\log \beta + \bar H_S + \log \frac{1-\mathrm{e}^{-\beta \exp(\bar H_S)}}{\beta \mathrm{e}^{\bar H_S}}\right] \right) \notag \\
  %  &= \sum_{\omega \in\Omega}\exp\left(-\sum_S \left[\log \beta + \bar H_S - \frac{\beta}{2}\mathrm{e}^{\bar H_S} + \cdots \right] \right).
%\end{align}

Physical predictions of statistical mechanics are unaffected by constant offsets to the Hamiltonian.  As such, given any observable $A:\Omega \rightarrow \mathbb{R}$ independent of the ``transmutation bosons", we can calculate its correlation functions at effective temperature $\beta_{\mathrm{eff}}=1$ using effective Hamiltonian \begin{equation}
    H_{\mathrm{eff}}(\beta) = \sum_S \left[\bar H_S + \frac{\beta}{2}\mathrm{e}^{\bar H_S}+ \cdots\right].
\end{equation}
At any finite $N$ (and finite $|\Omega|$) we conclude that \begin{equation}
    \lim_{\beta \rightarrow 0} \langle A\rangle_{\Omega_{\mathrm{new}}} = \lim_{\beta \rightarrow 0} \frac{\sum \mathrm{e}^{-H_{\mathrm{eff}}(\beta,\omega)}A(\omega)}{\sum \mathrm{e}^{-H_{\mathrm{eff}}(\beta,\omega)}} = \frac{\sum \mathrm{e}^{- \bar H} A}{\sum \mathrm{e}^{- \bar H}} \label{eq:limitbeta0}
\end{equation}
which is a correlation function at $\beta=1$ of $\bar H$.  In particular, if the original system was ordered at $\beta=1$ the new system with the bosons $n_S$ will be ordered at high temperature.

Strictly speaking, the derivation above relies on the fact that $N$ and $|\Omega|$ are finite.  We expect that even if one takes the thermodynamic limit \emph{first}, the conclusion holds so long as $\bar H$ is not infinitesimally close to a phase transition, but we will not attempt a proof of this conjecture here.  In the special case where the function $H_S$ is proportional to a product of $\pm1$-valued degrees of freedom, as in the Ising model, we can exactly evaluate $H_{\mathrm{eff}}$ at finite $\beta$ by using \eqref{eq:ZIsing} and \eqref{eq:ZIsing2} from the main text.

Indeed, combining the derivation above with the explicit model \eqref{eq:ZIsing} from the main text, we can clearly see that any classical phase of matter realized at any $\beta>0$ in a model with finite unconstrained configuration space, can be realized through entropic order.  This would include ferromagnetism as in the Ising model, $\mathrm{S}_n$-broken Potts model ``magnets", and any other more exotic phase the reader desires.

We can further ``improve" the construction \eqref{eq:A13} such that the limit $\beta \rightarrow 0$ reproduces arbitrarily low temperature for $\Omega$-correlation functions after ``integrating out" the transmutation bosons.  For simplicity let us focus on the case $\Omega=\mathbb{Z}_2^N$ to keep the discussion simpler.  Taking $H_S = -J_S \cdot s_S$ (where $s_S\in\lbrace\pm1\rbrace$ is a product of the $\mathbb{Z}_2$-valued variables), our strategy is to replace \eqref{eq:A13} with the yet more contrived Hamiltonian \begin{equation}
    H = \sum_S (n_S+1)^{\eta_S -\gamma_Ss_S} \label{eq:morecontrived}
\end{equation}
where we choose \begin{subequations}
    \begin{align}
        \eta_S &= \frac{\Lambda}{\Lambda^2-J_S^2}, \\
        \gamma_S &= \frac{J_S}{\Lambda^2-J_S^2},
    \end{align}
\end{subequations}
for any $\Lambda>\max(|J_S|)$.  The key observation is that as $\beta \rightarrow 0$, we have (for $s\in\lbrace\pm1\rbrace$) \begin{equation}
    \sum_{n=0}^\infty \exp\left(-\beta (n+1)^{\eta - \gamma s}\right)\approx \int\limits_0^\infty \mathrm{d}n\exp\left(-\beta n^{\eta - \gamma s}\right) = T^{(\eta+\gamma s)/(\eta^2-\gamma^2)}\Gamma\left(1+\frac{\eta+\gamma s}{\eta^2-\gamma^2}\right).
\end{equation}
Therefore when we sum over $n_S$ in the partition function, we find \begin{equation}
    Z(\beta) = \sum_s \exp\left[-\sum_S \left(\text{constant} + \log\frac{1}{\beta}\cdot J_Ss_S + \cdots \right)\right]
\end{equation}
where $\cdots$ denotes subleading terms in the limit $\beta \rightarrow 0$.  In this model, we see that the $\beta \rightarrow 0$ limit actually corresponds to the $\beta_{\mathrm{eff}}\rightarrow \infty$ limit of  the ``original" spin model. 

The construction \eqref{eq:morecontrived} has interesting implications for entropic-ordered spin glasses.  From a computational complexity perspective, spin glasses are very intriguing phases exhibiting a complex energy landscape with extensive barriers separating distinct low-energy regions of configuration space (``wells"): see \cite{Montanari_book} for an introduction.  Condensation of the measure $\mathbb{P}_\beta$ into such wells arises at sufficiently low temperature, heralding a spin glass phase.  In physics, one usually talks about ``replica symmetry breaking" when this condensation occurs, although we feel it is more natural to think about measure condensation directly, as discussed around \eqref{eq:condensation}.  One of the most ``practical" manifestations of spin glass order arises in the solution of NP-hard combinatorics problems, where the exponentially slow runtime of finding a solution to the combinatorial problem is closely related to the complex energy landscape of a glass \cite{krzakala}.  By coupling such a combinatorial problem, which can readily be expressed as an Ising-like model \cite{Lucas:2013ahy}, to ``transmutation bosons", we build a model which is in the glass phase as $\beta\rightarrow 0$. Moreover, if $\beta_{\mathrm{eff}} \rightarrow\infty$ as $\beta\rightarrow 0$, we can conclude that \emph{almost all states below energy $E$ are difficult to obtain as $E\rightarrow \infty$} with any known classical algorithm.  A randomly generated state, which will be very far from any solution to the combinatorial problem, has too few bosonic excitations -- it is an atypical state at its energy density.   Thus, in a precise sense (with the order of limits taken as written above), there exist models where ``almost all accessible states" are computationally hard to find.  As one explicit example, if we take the spin glass model to be the Sherrington-Kirkpatrick model, finding an arbitrarily low energy density state requires a diverging computational time scale as the (relative) energy density to the ground state vanishes:  see e.g. \cite{montanari19} for recent advances in efficient algorithms for this task.  We will return to the point that the high temperature Gibbs state may be computationally challenging to construct, in quantum mechanical models, in Appendix \ref{app:qm}.

As a final remark, let us also describe how to use transmutation bosons to build a model of entropic-ordered superfluidity at arbitrarily high temperatures.  Conventional superfluid phases of matter spontaneously break a U(1) global symmetry, and this can occur at finite temperature in $d=3$ spatial dimensions.  So, consider a cubic lattice with a $\mathrm{U}(1)$-valued degree of freedom $\theta_v \sim \theta_v+2\pi$ at each vertex $v$,\footnote{Although the configuration space $\Omega$ is now a manifold, because the manifold has finite volume for finite $N$, our general remarks about the existence of the limit $\beta \rightarrow 0$ in \eqref{eq:limitbeta0} continue to apply.} with XY Hamiltonian \begin{equation}
    \bar H = -J\sum_{u\sim v}\cos(\theta_u-\theta_v).
\end{equation}
The global U(1) symmetry correpsonds to $\theta_v \rightarrow \theta_v+c$ for some constant $c$.  In classical statistical mechanics, it is rigorously established that this model has long-range order for sufficiently large $J$, at $\beta_{\mathrm{eff}}=1$.  Upon adding a transmutation boson on every link, we obtain a model of entropic-ordered superfluidity as $\beta\rightarrow 0$.

\section{Kramers-Wannier duality}
For completeness we now make some comments about  Kramers-Wannier duality, which famously exchanges high and low temperature in the Ising model. 
A precise statement is that the partition function of the Ising$/\mathbb{Z}_2$ model is ``ordered'' at high temperature. However, this type of ``order'' is distinct from what we have done in our work for several reasons:

\begin{itemize}

\item Most importantly, this type of order cannot be understood in terms of the probability density collapsing into different clusters (related by symmetry), as defined in Appendix~\ref{app:classical}. This is because the partition function does not take the form of \eqref{eq:appendixZ}: the Ising$/\mathbb{Z}_2$ model is defined with a constraint on the holonomy of a $\mathbb{Z}_2$ gauge field, and if one tries to remove this constraint so that one can measure order with a local variable, the ``probability measure" ceases to be real-valued.

\item The physical meaning of this ``order'' can be understood already in the original Ising model as the insensitivity of the high-temperature phase to flipping some ferromagnetic couplings to anti-ferromagnetic ones. This does not reflect order in an experimentally measurable sense, rather it simply reflects the fact that the high temperature phase is insensitive to the spin-spin interaction. (Furthermore, when we discuss quantum mechanics in Appendix \ref{app:qm}, the ``order parameter" used to measure this high-temperature order will not be a physical observable on a Hilbert space.)
\end{itemize}

Let us now explain these points in some detail. 
The Ising$/\mathbb{Z}_2$ model, which is also known as Ising$^*$, is defined by the original Ising model coupled to a flat $\mathbb{Z}_2$ gauge field. The gauge field variables $U_{ij}\in \mathbb{Z}_2=\lbrace \pm 1\rbrace$ transform by left-right conjugation and we have the partition function 
\begin{equation}\label{part}Z=\sum_{\{U,S,\lambda\}}\mathrm{e}^{\sum_{(ij)} \beta J U_{ij}S_i S_j+\lambda_P\log \prod_{ij\in P} U_{ij}}~,\end{equation}
where $\lambda_P\in \{0,1\}$ is a new Ising-like variable which plays the role of a Lagrange multiplier, which imposes the flat connection condition once we sum over $\lambda_P$. Indeed, for those plaquettes which are flat $\log \prod_{ij\in P} U_{ij}=0$ and otherwise $\log \prod_{ij\in P} U_{ij}=i\pi$. Then if we perform the sum over $\lambda_P$, the gauge field is set to be flat.

The partition function~\eqref{part} has a $\mathbb{Z}_2$ symmetry which flips the values of $\lambda_P $ as $0\leftrightarrow 1$ (i.e. $\lambda_P\to 1-\lambda_P$). This is easy to see  since each edge is shared by two plaquettes the exponent has terms such as  $(\lambda_P+\lambda_{P'}) \log U$ and this term depends only on whether the two $\lambda$'s are equal or different and hence there is a symmetry in flipping all of them together. This is a $\mathbb{Z}_2$ global symmetry of the Ising$^*$ model. Sometimes it is referred to as ``quantum $\mathbb{Z}_2$ symmetry'' or the ``dual $\mathbb{Z}_2$ symmetry.'' 

%If there are edges on the boundary then we can put a boundary condition $\log U =1$ so that the symmetry remains, or we can choose other boundary conditions which would break the symmetry. 

 It is noteworthy that the term proportional to $\lambda_P$ is temperature independent and also not real. That is why we cannot interpret~\eqref{part} as an ordinary Gibbs ensemble.

 If we ignore global issues 
and sum over $\lambda_P$, we set the gauge field to be flat which allows us to set $U_{ij}=1$.  The model then becomes just the standard Ising model with coupling $\beta J$. Alternatively, this model can be solved by choosing the gauge $S_i=1$.\footnote{This is always possible, even on closed manifolds.}
Then, we can write the partition function as a sum over the $U,\lambda_P$ with the partition function\footnote{For simplicity, we have dropped an overall prefactor of $2^N$, where $N$ denotes the total number of vertices; this accounts for the sum over $S_i$.} 
\begin{equation}\label{fullP} Z=\sum_{\{U,\lambda\}}\mathrm{e}^{\sum_{(ij)} \beta J U_{ij}+\lambda_P\log \prod_{ij\in P} U_{ij}}~.\end{equation}
%So far the manipulations are valid in any topology. 
%This is because for every configuration of the spins we can set the gauge $S_i=1$ which amounts to transforming $U\to USS$ and this does not change the plaquette terms so we can just make a change of variables and call it the new $U$. 
%We just need to say that the map 
%$\tilde U= U SS$ is invertible, that is, for every $S$, every element $U$ maps to exactly one element $\tilde U$ and the map is unto. 
The sum over each link can be done separately. Then we get the partition function 
\begin{equation}\label{eq:lambdaZ}
    Z = \sum_{\{\lambda_P\}}\prod_{P\sim P^\prime} \left[\mathrm{e}^{\beta J}+\mathrm{e}^{-\beta J}\mathrm{e}^{\mathrm{i}\pi (\lambda_P+\lambda_{P'})}\right] .
\end{equation}
where $P\sim P^\prime$ denotes that the two plaquettes share an edge. We can think that $P$ lives on the dual lattice; the partition function above implies that $\lambda_P$ interact via nearest-neighbor interactions.
If $\lambda_P=\lambda_{P'} $ we get one and the same value while if they are different we get another value. Therefore, defining 
%So we can write in terms of dual variables $\tilde S=\pm1$ that live on the plaquettes such that  
\begin{subequations}
    \begin{align}
        e^{\beta J}+e^{-\beta J} &= \mathrm{e}^{A+\beta^\prime J^\prime}, \\
        \mathrm{e}^{\beta J}-\mathrm{e}^{-\beta J} &= \mathrm{e}^{A-\beta' J' },
    \end{align}
\end{subequations}we can re-write \eqref{eq:lambdaZ} as 
\begin{equation} \label{eq:lambdaZ2}
   Z = \sum_{\lbrace \lambda_P\rbrace } \mathrm{e}^{ \sum_{P\sim P^\prime} (A+\beta^\prime J^\prime\lambda_P\lambda_{P^\prime})} .
\end{equation}
This famously exchanges high and low temperatures with $\mathrm{e}^{2\beta'J'}=\coth(\beta J)$ the duality relation between temperatures.

The gauge invariant observables in our theory~\eqref{part} are insertions of $\lambda$, sometimes called the Kadanoff-Ceva disorder operator \cite{Fradkin:2016ksx}, such as:
\begin{equation}
\langle(1-2\lambda_{P})(1-2\lambda_{P'}) \rangle = \frac{1}{Z} \sum_{\lbrace U,\lambda\rbrace}\mathrm{e}^{\sum_{ij}\beta J U_{ij}+\sum_{P''}\lambda_{P''}\log\prod_{ij\in {P''}} U_{ij}}(1-2\lambda_P)(1-2\lambda_{P'})
\end{equation}
If we first sum over $U$, we have seen that the partition function \eqref{eq:lambdaZ2} is a sum over real-valued terms, which we can indeed interpret as a partition function, but at a new effective temperature.  If we instead try to think in terms of the original degrees of freedom, 
%including $U$, where $\beta$ represented inverse temperature, 
what happens is that the sum over $\lambda$s modifies the holonomy $\prod_{ij\in P} U_{ij} = 1$ to $\prod_{ij\in P} U_{ij} = -1$, around all plaquettes where $1-2\lambda$ has been inserted.

% In terms of the original theory, %which is obtained by recklessly solving for $U$, the interpretation is more interesting.
% the sum over $\lambda_P$ is now modified from $1+e^{\log\prod U}$ to be just 
% $e^{\log\prod U}$ -- it is more convenient to insert instead $\langle 1-2\lambda_P\rangle $ then the sum is modified to 
% $1-e^{\log\prod U}$, which projects on holonomy -1. The effect of the $\lambda$ insertions is therefore essentially to flip the holonomy constraint to be -1 on certain plaquettes. %essentially the effect of the $\lambda$ insertion. 

We can solve for the gauge field with the new constraint.  Away from the plaquettes $P$ and $P^\prime$ where we have flipped holonomies, the flatness condition on $U$ allows to set (ignoring global issues)  $U_{ij}=1$.
%\footnote{For readers who are unfamiliar with  this point, an analogy is as follows:  $U$ is a `discrete 1-form' on the square lattice triangulation of the plane. The flatness condition $\mathrm{d}U=0$ implies that $U=\mathrm{d}S$.  For $\mathbb{Z}_2$-valued degrees of freedom this means $U_{ij}=S_iS_j$.}  
At the plaquette $P$, we must choose some edge $i^\prime j^\prime \in P$ such that $U_{i'j'}=-1$ to obey the holonomy constraint.  We then pick an arbitrary path $\gamma$ from $P$ to $P'$ on the dual lattice, and along this path we set $U_{ij\in\gamma}=-1$.  The holonomy $\prod U$ then flips to $-1$ only at $P$ and $P'$.  One can show using gauge-invariance of the Ising model that the choice of path does not modify the partition function.  So now, we see that \begin{equation}
    \langle(1-2\lambda_{P})(1-2\lambda_{P'}) \rangle = \frac{1}{Z} \sum_{\lbrace S\rbrace}\mathrm{e}^{\sum_{ij \notin \gamma} \beta J S_i S_j - \sum_{ij\in\gamma}\beta JS_iS_j}
\end{equation}
where $Z$ continues to represent the original Ising model partition function, and the expression in the sum represents a modified partition function where the bonds along the path $\gamma$ are flipped from $J\rightarrow -J$.   Therefore, insertion of $\lambda$ amounts to flipping interactions in the Hamiltonian. 

The correlators of $\lambda_P$ exhibit ``order'' because at high temperature, the partition function is insensitive to whether bonds are ferromagnetic or antiferromagnetic: 
%in particular, $\langle \lambda_P\lambda_{P'}\rangle=1$ at $\beta=0$, corresponding to $\beta'=\infty$ in ``dual temperature".  
Again, we stress that $\lambda_P$ is not a valid observable within the framework of Appendix \ref{app:classical}, nor would it be valid within a quantum mechanical framework (Appendix \ref{app:qm}).

%\andy{old text here:} We choose an edge with $U=-1$ and then draw a path on the dual graph to infinity or to the boundary -- that path leads to $U=-1$ at every intersection with the original lattice (again, ignoring global issues). In terms of the original spin $S$ degrees of freedom, we are flipping the coupling constant $J_{ij}=J U_{ij}$ on every such edge. 
%By gauge transformations we can change the path at will but not the plaquette through which it starts.  
%Thus we have re-derived the familiar claim that the $\lambda_P$ are the disorder operators in the Ising model [].

%In summary, we see that $\lambda_P$ correlators exhibit ``order'' but the partition function is not in terms of a Gibbs ensemble~\eqref{fullP} and if the $\lambda_P$ are integrated out, then the ``order'' corresponds to the response to flipping the couplings from ferromagnetic to anti-ferromagnetic along a half-infinite line. There is no measure condensation in $\mathbb{P}_\beta$.

%A more subtle non-example of high-temperature order is the ``disorder operator" in the two-dimensional Ising model, which is ordered at high temperatures, but not a valid observable!  To avoid potential confusion, we will discuss this non-example (and why it is a non-example) of high-temperature order in some detail.  

%\andy{to be continued, can cut/paste some of zohar's stuff here}

\section{Markov chain simulations for the classical bosonic lattice gas}

In this appendix, we detail how to simulate the classical bosonic lattice gas \eqref{eq:numericmodel} on a two-dimensional $L\times L$ square lattice with open boundary conditions, using a standard Metropolis algorithm \cite{Metropolis1953,Hastings1970} for sampling the Gibbs distribution \eqref{eq:thermalensemble}.  

First let us very briefly review how to perform Markov chain simulations. At time $t=0$, we pick an initial microstate $\mathbf{n}$.  The goal is to construct a transition matrix $\mathsf{P}(\mathbf{n}\rightarrow \mathbf{n}^\prime)$, with $\mathsf{P}^2(\mathbf{n}\rightarrow \mathbf{n}^{\prime\prime}) = \sum_{\mathbf{n}^\prime}\mathsf{P}(\mathbf{n}\rightarrow \mathbf{n}^\prime)\mathsf{P}(\mathbf{n}^\prime\rightarrow \mathbf{n}^{\prime\prime})$, such that 
\begin{equation}
   \lim_{\tau\rightarrow\infty} \mathsf{P}^\tau(\mathbf{n}\rightarrow \mathbf{n}^\prime) = \frac{\mathrm{e}^{-\beta H(\mathbf{n}^\prime)}}{Z(\beta)}. \label{eq:MCconvergence}
\end{equation}
Established theorems \cite{levin_markovchains} show that the limit \eqref{eq:MCconvergence} is unique if for sufficiently large $\tau$, $\mathsf{P}^\tau(\mathbf{n}\rightarrow \mathbf{n}^\prime)>0$ for all $\mathbf{n}$ and $\mathbf{n}^\prime$.  It is generally quite easy to obtain this condition. Slightly more non-trivial is to assure that the chain $\mathsf{P}$ has the correct stationary distribution on the right hand side of \eqref{eq:MCconvergence}.  This can be established by demonstrating the \emph{detailed balance} (or, time-reversal symmetry) condition: 
\begin{equation}
    \mathsf{P}(\mathbf{n}\rightarrow \mathbf{n}^\prime)\mathrm{e}^{-\beta H(\mathbf{n})} = \mathsf{P}(\mathbf{n}^\prime\rightarrow \mathbf{n})\mathrm{e}^{-\beta H(\mathbf{n}^\prime)}. \label{eq:detailedbalance}
\end{equation}
Granted that the limit on the left hand side of \eqref{eq:MCconvergence} exists, \eqref{eq:detailedbalance} ensures that the limit equals the right hand side.   We numerically evaluate expectation values of local observables in this thermal ensemble by evaluating: \begin{equation}
    \langle A\rangle = \sum_{\mathbf{n}}\mathbb{P}_\beta(\mathbf{n})A(\mathbf{n})\approx \frac{1}{t_1-t_0}\sum_{t=t_0}^{t_1-1}A(\mathbf{n}_t)
\end{equation}
where $\mathbf{n}_t$ is a random variable corresponding to the microstate that the Markov chain is in at time $t$.

To handle the infinite local confiugration space $\Omega = (\mathbb{Z}_{\ge 0})^{L^2}$ of the model \eqref{eq:numericmodel}, we introduce the following transition matrix: \begin{equation}
    \mathsf{P}(\mathbf{n}\rightarrow \mathbf{n}^\prime) = \left\lbrace \begin{array}{ll} 0 &\ \mathsf{d}(\mathbf{n},\mathbf{n}^\prime) > 1 \\ \displaystyle  \frac{1}{2L^2} \min\left(1 , \mathrm{e}^{-\beta [H(\mathbf{n}^\prime)-H(\mathbf{n})]}\right) p(|n_v-n_v^\prime|) &\ \mathsf{d}(\mathbf{n},\mathbf{n}^\prime) = 1 \text{ and } n_v^\prime \ne n_v \\ \displaystyle 1 - \sum_{\mathbf{n}^\prime \ne \mathbf{n}}\mathsf{P}(\mathbf{n}\rightarrow \mathbf{n}^\prime) &\ \mathbf{n}^\prime = \mathbf{n}\end{array}\right., \label{eq:markovchainrules}
\end{equation} 
where we recall the definition of Hamming distance $\mathsf{d}$ from Appendix \ref{app:classical}, and we have defined    \begin{equation}
        p(k) = \mathrm{e}^{-\beta(k-1)}(1-e^{-\beta}).
    \end{equation}
    We can interpret this Markov chain with the following intuitive algorithm.  For each time step:
\begin{itemize}
    \item uniformly at random choose a vertex $v$ out of $L^2$ possibilities,
    \item decide whether to attempt to increase or decrease the number of particles on $v$; the choice is made with probability $\frac{1}{2}$ for each. Then, attempt to replace $n_v$ with $n_{v^\prime} = n_v \pm k$ with probability $p(k)$.  Do not attempt to decrease the number of particles if $k>n_v$.
    \item Accept the potential change $n_v \rightarrow n_v^\prime$ with ``Metropolis" probability $\min(1,\mathrm{e}^{-\beta\Delta E})$, namely always accept an energy-lowering move, and sometimes accept an energy-raising move.
\end{itemize}
To see that \eqref{eq:MCconvergence} holds, notice that we may take $\tau=L^2$ and find a non-zero probability $\mathsf{P}^\tau(\mathbf{n}\rightarrow \mathbf{n}^\prime)$ by first changing $n_1\rightarrow n_1^\prime$, then $n_2\rightarrow n_2^\prime$, and so on.  It is manifest from the form of \eqref{eq:markovchainrules} that \eqref{eq:detailedbalance} holds, so we conclude that $\mathsf{P}$ samples from the correct distribution.  It is \emph{not} guaranteed that this algorithm is computationally efficient, and indeed we emphasize that much more efficient algorithms are known for sampling spin models such as the Ising model in very large system sizes, as discussed in \cite{Swendsen1987MC,Wolff1989MC}.  But we will find our crude algorithm satisfactory for establishing entropic order and the universality class of the phase transition.  

To run the simulation, we must also choose an initial microstate $\mathbf{n}$. We typically do this by either randomly selecting $0\le n_v \le T$ on each lattice site, \emph{or} by restricting the occupied sites to those on a specific sublattice of the bipartite square lattice.

%To get an intuitive understanding of the typical configurations of particles, we take snapshots of the lattice at late times at different temperatures. As shown in Fig.\ref{fig:SSB_config}, at high temperature ($\beta=0.05$), the configuration exhibits an ordered checkerboard pattern where every other site is occupied by a large number of particles; At low temperature ($\beta=0.5$), large occupation is discouraged and hence most of the sites are either empty or occupied by a small number of particles of order $O(1)$.

% \begin{figure}[!t]
%     \centering
%     \includegraphics[width=0.8\linewidth]{Boson_data_collapse.pdf}
%     \caption{The data collapse of the order parameter $\widetilde{\Delta}$. $\beta=1/8$ (not to be confused with the inverse temperature) and $\nu=1$ are the critical exponents in the $2d$ Ising model. We take the interaction strength $U=1$ and $a=2$, which gives the critical point $T_{\mathrm{c}}=1/\beta_c\approx1/0.19$.}
%     \label{fig:boson_data_collapse}
% \end{figure}

As we summarized in the main text, it is often most natural to detect the thermodynamic phase transition by a local order parameter.  In a classical lattice gas on the square lattice, such an order parameter is well-known \cite{} to be the difference of the particle density on the A vs. B sublattices.  We square this quantity to ensure that when sampling over many realizations of the model, we are insensitive to which of the two sublattices the particles cluster on.  Therefore we investigate
\begin{equation}
    \Delta(\beta) = \frac{\langle (N_{\mathrm{A}}-N_{\mathrm{B}})^2\rangle}{L^4},
\end{equation}
with $\langle \cdots \rangle$ determined numerically as described above.\footnote{Note that in a high temperature ordered phase, we do not expect the Markov chain to actually traverse from microstates with $N_A>N_B$  to those with $N_B>N_A$, because the mixing time to do so is expected to diverge as $\exp(\Theta(L))$ \cite{Thomas1989}.  However, $\Delta(\beta)$ takes the same value in either of the two clusters in the ordered phase, so we can accurately deduce $\Delta(\beta)$ well before this mixing time.}
To verify that our numerics are accurate, we have a few simple heuristic benchmarks.  As stressed in the main text, as $\beta \rightarrow 0$ the entropic penalty from straying from the checkerboard states is extremely large, suggesting that an average site on the A sublattice will see all four of its neighboring sites unoccupied.  In this case, we can easily evaluate the average occupation number on that site:  
\begin{equation}
    \langle n_{v\in \mathrm{A}}\rangle=\dfrac{\displaystyle \sum_{n=0}^\infty n\mathrm{e}^{-\beta n}}{\displaystyle \sum_{n=0}^\infty \mathrm{e}^{-\beta n}}= \frac{1}{1-\mathrm{e}^{-\beta}}.
\end{equation}
Due to the central limit theorem, we can estimate that 
\begin{equation}
    \Delta\approx\frac{\langle N_{\mathrm{A}}^2\rangle}{L^4}\approx\frac{\langle N_A\rangle^2}{L^4}\approx \frac{1}{L^4}\left(\frac{L^2}{2}\langle n_{x\in\mathrm{A}}\rangle\right)^2=\frac{1}{4(1-\mathrm{e}^{-\beta})^2}.
\end{equation}
This estimate should be off by an O(1) factor which vanishes as $\beta \rightarrow 0$ (we do not attempt to estimate this factor).  
This heuristic is verified in Fig.~\ref{fig:boson}b. As $\beta$ increases, $\Delta$ decreases until the curves for different system sizes cross at $\beta_c\approx 0.19$, above which they start to deviate.  This is our numerical sign that we enter the ordered phase.

To show that this phase transition belongs to the Ising universality class, we consider the following
order parameter:
\begin{equation}
    \widetilde{\Delta} = \frac{\langle(\sigma_{\mathrm{A}}-\sigma_{\mathrm{B}})^2\rangle}{L^4} = \left\langle \left(\sum_{x,y=1}^L (-1)^{n_x+n_y} (1-\delta_{n_{xy},0})\right)^2 \right\rangle.
\end{equation}
Notice that $\widetilde{\Delta}>0$ is actually a stronger condition than $\Delta>0$, as we forbid even a single particle (rather than a stack of them) from existing on the ``wrong" sublattice.  As is standard, we expect corrections to critical phenomena due to finite size \cite{Fernandes_2007}: namely we expect:
\begin{equation}
    \widetilde{\Delta}=L^{-2\bar\beta/\nu}f\left[(T-T_{\mathrm{c}})L^{1/\nu}\right], \label{eq:finitesizescaling}
\end{equation}
where $\bar\beta=1/8$ and $\nu=1$ are the critical exponents of the $2d$ Ising model and $f$ is a model-specific function. Fig.~\ref{fig:boson}c confirms that at different system sizes, the numerically measured $\widetilde{\Delta}$ is indeed consistent with the form \eqref{eq:finitesizescaling} near $T_{\mathrm{c}}$.  Therefore,  despite the classical bosonic lattice gas having $\mathbb{Z}_2$ SSB at high temperature, rather than low temperature, the transition itself is consistent with the Ising universality class.

\section{Quantum lattice models}\label{app:qm}
In this appendix, we now discuss quantum mechanical lattice models.  We will, to the extent possible, ensure that the development closely mirrors Appendix \ref{app:classical}.  As there, we will only take the thermodynamic limit at the end of a calculation to simplify the formalism.

In quantum mechanics, we define a Hilbert space $\mathcal{H} = \mathrm{L}^2(\Omega)$ of normalizable wave functions over a configuration space $\Omega$.  Here we will always take $\Omega$ to be countable, for simplicity.  A Hamiltonian $H$ corresponds to a Hermitian linear operator on $\mathcal{H}$ whose eigenvalue spectrum is bounded from below.   The Gibbs state is defined as a mixed state (density matrix) \begin{equation}
    \rho_\beta := \frac{\mathrm{e}^{-\beta H}}{Z(\beta)} \label{eq:quantumgibbs}
\end{equation}
where \begin{equation}
    Z(\beta) = \mathrm{tr}\left(\mathrm{e}^{-\beta H}\right).
\end{equation}
As in Appendix \ref{app:classical}, we assume that $Z(\beta)$ is well-defined for any $\beta>0$.
%a necessary and sufficient condition for this is that $H$ has a discrete spectrum (with finite-dimensional eigenspaces).

Notice that \emph{all} classical models within the scope of formalism from Appendix \ref{app:classical} can be ``lifted" to quantum models, by simply choosing $H$ to be a diagonal matrix in the basis $|\omega\rangle$ for $\mathcal{H}$. 

It has recently been shown \cite{Bakshi:2024cqr} that under similar assumptions to Appendix \ref{app:classical} -- namely, $H$ is $k$-local, $\mathcal{H}$ is a tensor product of finite-dimensional Hilbert spaces (analogous to the assumption \eqref{eq:unconstrainedspace} on $\Omega$), and $H$ is a sum of bounded terms, finitely many of which act on a given qubit -- $\rho_\beta$ is separable above a critical temperature.  For $\beta<\beta_*=\mathrm{O}(1)$, we may write \begin{equation}
     \rho_\beta = \sum_a p_a |\psi_a\rangle\langle \psi_a| \label{eq:separable}
 \end{equation}
 where $|\psi_a\rangle$ is a product state and $p_a$ is some probability distribution with $0<p_a<1$ and $\sum_a p_a=1$, ensuring $\mathrm{tr}(\rho_\beta)=1$.  This has been coined a ``heat death of entanglement" at high temperatures in quantum systems.  A less stringent statement, namely the absence of long-range order at high temperature, has also been known for quantum systems \cite{kliesch2014locality,Yin:2023zfq}, including certain Bose-Hubbard type models of interacting bosons \cite{tong2025locallyinteractinglatticebosons}.

Intuitively, we might expect that at high temperatures, most physical systems admit an effectively classical description. 
 As highlighted in the main text, however, we can use entropic order to build a phase of matter which is fundamentally quantum at high temperatures, such as a phase with entropic topological order, which will violate the no-go theorem on entanglement at high temperatures.  As before, we will use unbounded Hilbert spaces to achieve this result.

 Let us briefly review what topological order is in lattice models with a finite-dimensional Hilbert space $\mathcal{H}$.  We say that  a density matrix $\rho$ has topological order if and only if for any separable density matrix $\rho_{\mathrm{sep}}$, as in \eqref{eq:separable}, and finite-depth $k$-local quantum channel $\mathcal{E}$, \begin{equation}
     \lVert\rho - \mathcal{E}(\rho_{\mathrm{sep}}) \rVert_1 \ge \epsilon;
 \end{equation}
 here $\lVert A\rVert_1$ denotes the sum of absolute values of $A$'s eigenvalues, for any O(1) $\epsilon$ independent of system size \cite{finiteTtopological}.   Intuitively, this definition of topological order implies that the state is so deeply entangled that no finite circuit can generate the entanglement.\footnote{Note that we cannot write $\lVert \mathcal{E}(\rho)-\rho_{\mathrm{sep}}\rVert_1\ge\epsilon$, as one can e.g. trace out every degree of freedom in $\rho$ to get the identity matrix, which is separable.}

 Perhaps the simplest and most important manifestation of quantum topological order is in a CSS quantum error-correcting code \cite{Calderbank_1996,Steane_1996}, which can be intuitively understood as a Hamiltonian of the form \begin{equation}
     H = -\sum_{C_X}\prod_{i\in C_X}X_i -\sum_{C_Z}\prod_{i\in C_Z}Z_i = -\sum_{\text{stabilizer }S}S \label{eq:CSS}
 \end{equation}
 on Hilbert space $\mathcal{H} = (\mathbb{C}^2)^{\otimes n}$ of $n$ qubits.  Here $C_X$ and $C_Z$ represent $X$-type and $Z$-type parity checks respectively.  All of the Pauli strings in the above Hamiltonian commute, such that the spectrum of $H$ can be found exactly, as in a classical model.  However, because we have products of Pauli $X$ and $Z$ which do not commute on-site, the eigenstates of $H$ are highly entangled.  In particular, we may be able to find $k$ \emph{logical operators}, Pauli strings of $X$-type or $Z$-type which commute with every single parity check.  There are then $2^k$ ground states of $H$ which simultaneously satisfy all parity checks, but encode different logical qubits.   The code distance $d$ is the fewest number of sites on which a logical operator acts.   %\andy{I can give a more formal description, but maybe for our readers it is pointless}

It is known that there are certain codes whose Gibbs states \eqref{eq:quantumgibbs} are topologically ordered at large but finite $\beta$.  Examples include the four-dimensional toric code \cite{alicki2008thermal}, as well as certain codes which have the ``NLTS" property \cite{Anshu_2023}, implying that every single low-energy pure state (not just the Gibbs state) is topologically ordered.

We can use the exact same ``transmutation bosons" from Appendix \ref{app:classical} to realize this quantum topological order at arbitrarily high temperature.  Suppose that when $\beta=J>1$, the Gibbs state of \eqref{eq:CSS} is topologically ordered.   Then define a Hilbert space \begin{equation}
    \mathcal{H} := \mathrm{L}^2\left( \left(\mathbb{C}^2\right)^{\otimes n} \otimes \bigotimes_S \mathbb{Z}_{\ge 0} \right)
\end{equation}along with a Hamiltonian \begin{equation}
    H = \sum_S \left(a-bS \right)(n_S+1)
\end{equation}
where the boson operators $n_S$ commute with each other: $[n_S,n_{S^\prime}]=0$, and with the code stabilizers: $[n_S, S^\prime]=0$.\footnote{We can write $n_S$ as a quantum operator in terms of raising/lowering operators, but it is not necessary for this calculation to do so.} Because all operators in the density matrix commute, we can exactly evaluate the reduced density matrix on the qubits, \begin{equation}
    \mathrm{tr}_{n_S} \rho_\beta = \rho^{\mathrm{CSS}}\left(\beta_{\mathrm{eff}} = B(\beta)\right), \label{eq:reducedBbeta} 
\end{equation}
and it is identical to the thermal state of \eqref{eq:CSS} at temperature $B(\beta)$ given in \eqref{eq:Beq}.  If we choose the parameters $a$ and $b$ such that $B(0)$ lies in a regime of topological order, we obtain a model of high-temperature entropic topological order.  This is because it is not possible to trace out degrees of freedom in a topologically trivial density matrix to obtain a topologically ordered density matrix \cite{finiteTtopological}, and we have seen from \eqref{eq:reducedBbeta} that a certain reduced density matrix in the system has topological order.

We can make analogous comments to Appendix \ref{app:classicalexamples} regarding entropic ordered spin glasses.  The presence of entropic topological order implies that there exist quantum mechanical systems, nearly all of whose high-energy states are impossible to prepare using any finite-depth quantum channel.  The topological order carried in the qubit degrees of freedom, which carries low entropy, is outweighed by the high entropy of the fluctuating bosons $n_S$.  These represent clear counterexamples to a ``heat death of entanglement" at high temperature.

 We also note that \cite{finiteTtopological} did show that it is possible to have topological order at $T>0$ but not at $T=0$; our construction above points out that topological order can exist as $T\rightarrow\infty$.   \cite{hamma} also discussed trying to use bosonic degrees of freedom to stabilize topological order.

\section{Quantum field theory}
In this appendix, we provide the details of calculations of entropic order in quantum field theory. 

\subsection{$\mathrm{O}(N)$ models}
Let us consider the case of field theory in $2<d<3$ and show that in the model we also have an entropic order. We start with the action \eqref{eq:QFTL}, and then introduce Hubbard-Stratonovich field $\sigma$ such that
\begin{gather}
    S = \int \mathrm{d}^{d+1} x \left[\frac12 (\partial_\mu \phi_i)^2 + \frac12 \left(\partial_\mu \psi\right)^2 + \frac{1}{2} \sigma\left(\sum_i \phi_i^2 - \psi^2\right) + \frac{N \sigma^2}{4\lambda}\right] \label{eq:on_theory},
\end{gather}
In the strong coupling regime, we can neglect the quadratic term $\sigma^2$, which is irrelevant at the critical point.  Now let us compute the effective potential in the large $N$ limit around a configuration with uniform $\sigma = \bar\sigma$, $\psi = \sqrt{N}\bar\psi $, and $\phi^a=(\sqrt{N}\bar \phi,0,\ldots,0)$.  Integrating out $N-1$ bosons $\phi^a$ we find, at leading order in large $N$, the effective free energy density \begin{equation}
   \beta  \mathcal{F}(\bar \sigma,\bar\phi,\bar\psi)  =  \frac{N}{2}\left( \left[\log \left(-\Delta+\bar\sigma\right)\right]_{xx}  + \bar\sigma \left(\bar\phi^2-\bar\psi^2\right)\right) + \mathcal{O}(1) \label{eq:FON}
\end{equation}
For simplicity in what follows we neglect the overbar around the ``vacuum" of the fields.  The saddle point of the free energy obeys:
\begin{subequations}
\begin{align}
    \frac{1}{-\Delta+\sigma} &= \psi^2 - \phi^2 \label{eq:sigmaeqON} %\\
%    \sigma \psi &= \sigma \phi = 0
\end{align}
\end{subequations}
Note that we can only meaningfully evaluate $F$ at a saddle point of $\sigma$, which we will do henceforth: 
\begin{gather}
 \sigma = \sigma_*(\psi, \phi, T): \quad \frac{\partial \mathcal{F}}{\partial \sigma} (\sigma_*, \psi, \phi, T) = 0.
\end{gather} 
We will consider the configurations where $\sigma \ll T^2$.  Then we expand the propagator at coinciding points as
\begin{gather}
     \frac{1}{-\Delta+\sigma}  \approx \frac{\Gamma\left(\frac{2-d}{2}\right)  }{(4\pi)^\frac{d}{2}}  T \sigma^\frac{d-2}{2} + \frac{\pi^\frac{d-4}{2}}{4} \Gamma\left(\frac{2-d}{2}\right)\zeta(2-d) T^{d-1}.
\end{gather}
Solving \eqref{eq:sigmaeqON} we find that 
\begin{gather}
    \sigma_*^\frac{d-2}{2} = \frac{(4\pi)^\frac{d}{2}}{\Gamma\left(\frac{2-d}{2}\right) T} \left( \psi^2 - \phi^2 - \frac{\pi^\frac{d-4}{2}}{4} \Gamma\left(\frac{2-d}{2}\right)\zeta(2-d) T^{d-1}\right) \label{eq:sigma*}
\end{gather}
Now plugging in \eqref{eq:sigma*} into \eqref{eq:FON} we find %\begin{equation}
%    \mathcal{F}(\psi, \phi) = - \alpha T^{d+1}+\left(1-\frac{2}{d}\right) \left(\frac{(4\pi)^\frac{d}{2}}{\Gamma\left(\frac{2-d}{2}\right) T}\right)^{\frac{2}{d-2}} \left( \psi^2 - \phi^2 - \frac{\pi^\frac{d-4}{2}}{4} \Gamma\left(\frac{2-d}{2}\right)\zeta(2-d) T^{d-1}\right)^{\frac{d}{d-2}}
%\end{equation}
\begin{equation}
    \mathcal{F}(\psi, \phi) = -\frac{\,\Gamma(d)\,\zeta(d+1)}{2^{d-1}\,\pi^{d/2}\,\Gamma\left(\frac{d}{2}\right)} T^{d+1}+ \frac{2^{1-d} \Gamma\left(2 - \frac{d}{2}\right) T}{d \pi^\frac{d}{2}} \left( \frac{2^ d \pi^\frac{d}{2}}{\Gamma\left(1 - \frac{d}{2}\right) T } \left(\psi^2 - \phi^2\right) - (2\pi T )^{d-2} \zeta(2-d) \right)^\frac{d}{d-2}
\end{equation}
%\andy{I changed this from Fedor's version, plz check.  This version now has correct units. Also should we just calculate the constant $\alpha$ for completeness...?}  
This leads to \eqref{eq:mainfreeenergydensity} in the main text.
% \begin{gather}
%     F[\psi, \phi] = \frac12 \sigma (\phi^2 - \psi^2) + \frac{\pi^\frac{d-5}{2}}{8} \Gamma\left(\frac{3-d}{2}\right)\zeta(3-d) T^{d-2} \sigma -  \frac{\Gamma\left(\frac{1-d}{2}\right) T \sigma^\frac{d-1}{2} }{2(4\pi)^\frac{d-1}{2}}, \notag
%       F[\psi, \phi] = \frac{3-d}{d-1} \frac{2\Gamma\left(\frac{3-d}{2}\right) T \sigma_*^\frac{d-3}{2} }{(4\pi)^\frac{d-1}{2}}
% \end{gather}
%\andy{the 2 halves of this formula are not consistent with each other?} 

To make a connection to the idea of entropic order, we will try to find how many states with given energy $E$ and given order parameter $\psi$ are present here. That accounts to computation of the entropy $S[E,\psi]$ as a function of energy and order parameter $\psi$.  Performing the sum over discrete Matsubara frequencies explicitly in \eqref{eq:FON}: 
%First, we introduce some currents to our model, that will set $\psi$ to some given value. Since we work in the thermodynamic limit, this could be done in the following way; it accounts to fixing $\psi$ and not imposing the equations of motion for $\psi$. The same we will do for the field $\sigma$, but we will impose equations of motion for the field $\phi$ and set them to zero $\phi_i = 0$. From this we can find the free energy as a function of $\sigma, \psi$ and temperature.
33\begin{gather}
    \mathcal{F}[\sigma, \psi, T] = T \int \frac{\mathrm{d}^{d} p}{(2\pi)^{d}} \log \left[ 1 - \mathrm{e}^{-\frac{\sqrt{p^2 + \sigma}}{T}}\right] 
    +\frac12 \int \frac{\mathrm{d}^{d} p}{(2\pi)^{d}} \left(\sqrt{p^2 + \sigma} - p\right) - \frac{1}{2} \sigma \psi^2,
\end{gather}
This equation has a very simple physical meaning: the first term computes the free energy of the scalar fields  with mass $\sigma$, the second part computes the shift in the ground state (zero-point) energy of the free scalar fields due to the presence of non-zero $\sigma$ field, and the last part is the term responsible for interaction of the field $\sigma$ with the order parameter $\psi^2$. The energy density of such configuration is given by
\begin{figure}
    \centering
    \includegraphics[width=0.5\linewidth]{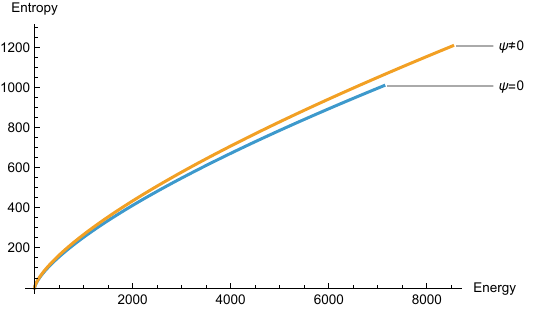}
    \caption{The plot of $\mathcal{S}(\psi, \mathcal{E})$ for $\psi = 0$ and $\psi \neq 0$.}
    \label{fig:entropic_order_on}
\end{figure}
\begin{gather}
    \mathcal{E}[\sigma, \psi, T] = \int \frac{\mathrm{d}^{d-1} p}{(2\pi)^{d-1}} \frac{\sqrt{p^2 + \sigma}}{\mathrm{e}^{\frac{\sqrt{p^2 + \sigma}}{T}} - 1}+ \frac12 \int \frac{\mathrm{d}^{d-1} p}{(2\pi)^{d-1}} \left(\sqrt{p^2 + \sigma} - p\right) - \frac{1}{2} \sigma \psi^2,
\end{gather}
again the first part is just an average energy of particles with mass $m$ at temperature $T$. And the entropy density of such configuration could be computed in the following way
\begin{gather}
    \mathcal{S}[\sigma, \psi, T] = \frac{\mathcal{E} - \mathcal{F}}{T}
\end{gather}
Now we fix $\psi, T$ and try to find the typical contribution that contributes to the given state, after plugging in for $\sigma=\sigma_*$ as before.
Then we can change $T$ and find entropy and energy $\left(\mathcal{E}(\psi, T), \mathcal{S}(\psi,T)\right)$ as a function of $\psi$ and $T$. Now if we invert the second equation and find $T$ as a function of order parameter $\psi$ and entropy $E$, that will allow us to find entropy as a function of order parameter $\psi$ and energy $E$.
That produces the plots in Figure \ref{fig:entropic_order_on}, which confirm that most of the high energy states are in the ordered phase, so the QFT indeed as entropic order. 

%\XYH{I add a numeric plot for $3+1d$ in my note. I first extremize $F$ over $\sigma$ then plot $F[\psi,T] = F[\sigma_*,\psi,T]$ against $\psi^2$ at a fixed $T$. It ceases to have a real positive $\sigma_*$ when $\psi^2 \geq T^2/12$, consistent with the analytic results. Moreover, $F[\psi,T]$ is minimized at $\psi^2 = T^2/12$. Similarly, the entropy $S[\psi,T]$ is maximized at the same point.   }

In $d=2$ we can find the explicit form of the free energy functional as a function of $\sigma,\psi$ but we should take into account the existence of $\psi^6$ potential \cite{Komargodski:2024zmt}. We get
\begin{gather}
    \mathcal{F}[\sigma, \psi] = - \frac{N}{4\pi \beta^3}\left(\frac13 \left(\sigma \beta^2\right)^\frac32 + 2 \sqrt{\sigma} \beta \operatorname{Li}_2\left(e^{-\sqrt{\sigma}\beta}\right) + 2 \operatorname{Li}_3 \left(e^{-\sqrt{\sigma} \beta}\right)\right) - \frac12 \sigma \psi^2 + \frac{g}{6!}\psi^6,
\end{gather}
due to the presence of the sextic interaction, the qualitative picture of \eqref{fig:entropic_order_on} is different from fractional dimensions case. Thus, if we fix some $\psi \neq 0$ and examine its entropy, we find that the line corresponding to $\psi = 0$ prevails at a certain energy $E_\psi$. However, for any energy $E$, there always exists some $\psi \neq 0$ with higher entropy than the disordered state. To be more precise we can check that the saddle point equations are
\begin{subequations}\begin{align}
    \psi^2 &= 2 \log\left(2 \sinh \frac{\beta \sqrt{\sigma}}{2} \right),  \\
     \sigma &= \frac{g}{5!}\psi^4,
    \end{align}
\end{subequations}
from which we can immediately see that $\psi \neq 0$.

%\andy{I think we only keep this part if we explicitly discuss the correction to the 1-loop effective potential from integrating out fluctuating photons.  otherwise we should just cut this.} 
We can continue this model to higher dimensions and even study the case of continuous groups being spontaneously broken, but the naive theory \eqref{eq:on_theory} is not UV complete.  Thus we have some high temperature $T_*$, above which we can no longer use this theory. Nonetheless, we would have a spontaneous symmetry breaking for arbitrary high temperature $T < T_*$. A more interesting model involves the addition of a gauge field for $\mathrm{U}(1)$ symmetry; this then represents a toy model for a superconductor (a microscopic model for which we will shortly discuss).   We consider
\begin{gather}
    \mathcal{L} = -\frac{1}{4e_0^2} F_{\mu\nu}^2 + \left|\partial_\mu \psi - i A_\mu\psi\right|^2 
    + \frac{1}{2}\left(\partial \phi_i\right)^2 + \frac{\lambda}{4N} \left(|\psi|^2 - \phi_i^2\right)^2 
\end{gather}
We again use the standard trick and introduce a Hubbard-Stratonovich field $\sigma$ and integrate all but $\phi = \phi_1$ field, after which we get
\begin{gather}
    \mathcal{F}[\psi, \phi, \sigma] = \frac{N}{2} \left[\log \left(-\Delta + \sigma\right)\right]_{xx} 
    +N\frac{\sigma^2}{4 \lambda_0 } + \frac12 \sigma \left(\phi^2 -  \left|\psi\right|^2\right) 
    +\frac12 (\partial_\mu \phi)^2 +  \left|\partial_\mu \psi - i A_\mu\psi\right|^2   - \frac{1}{4e_0^2} F_{\mu\nu}^2 . 
\end{gather}
Now we want to find a saddle point for this free energy in the large $N$ limit. The most natural ansatz is to set $A_0$ to be constant and say that the condensate of $\phi,\psi$ is of the order $\mathcal{O}(\sqrt{N})$. After proper rescaling we find 
\begin{gather} 
 \frac{1}{N}\mathcal{F}[\psi, \phi, \sigma] =  \frac{1}{2} \left[\log \left(-\Delta + \sigma\right)\right]_{xx} + \frac{\sigma^2}{4\lambda_0}  
    + \frac{1}{2} \sigma \left(\phi^2 - \left|\psi\right|^2\right) +  A_0^2 \left|\psi\right|^2
\end{gather}
We want first to consider the saddle point for \(\sigma\) that is given now by the following equations
\begin{gather}
    \frac{1}{-\Delta + \sigma} = \left|\psi\right|^2 - \phi^2 + \frac{\sigma}{\lambda_0}, \quad 
    \sigma \phi =  0, \quad \left(\sigma - A^2_0\right) \psi = 0, \quad A_0 \left|\psi\right|^2 = 0
\end{gather}
There are two divergences that could be absorbed in the renormalization of the mass of the fields and the coupling constant \(\lambda\). Thus we arrive at the following gap equation
\begin{gather}
    \frac{1}{-\Delta + \sigma} 
    = \int \frac{k^2 \mathrm{d}k}{2\pi^2} \frac{1}{\sqrt{k^2+\sigma}} \frac{1}{\mathrm{e}^{\beta\sqrt{k^2+\sigma}} - 1}  - \frac{1}{4}\sigma \log \frac{\sigma}{T^2_*}
\end{gather}
First consider the following solution with \(\psi \neq 0\). That yields \(\sigma = A_0 = 0\) and we get the following equation
\begin{gather}
 \frac{1}{12\beta^2} = \left|\psi\right|^2 - \phi^2, \label{eq:u1model_gap_equation}
\end{gather}
so we would have again the manifold vacua, which could be resolved by taking into account the $\frac{1}{N}$ correction, noting that all these vacua have $\psi \neq 0$. One can check that the fluctuations of the gauge fields are suppressed in the large $N$ limit and the equation \eqref{eq:u1model_gap_equation} holds for any temperature $T<T_*$ for sufficiently large $N$.  This implies that, at least in principle, entropic ordered superconductivity is possible.

\subsection{Superconductivity}
We now present a more microscopic model for a superconductor with entropic order.  Let us first briefly review the BCS theory for superconductivity, which we will modify in a simple way to realize entropic order. Consider a fermionic Hamiltonian with an attractive interaction 
%\Zohar{This is not just back-to-back scattering, should we restrict to back-to-back scattering for simplicity?}\XYH{I think this is the standard $H$ for BCS theory, and we might not need to worry about forward scatterings if we choose the mean-field ansatz as SC.}
\begin{align}
    H = \int \ud^d x ~\left[ \sum_{\sigma,a}c^\dagger_{\sigma,a}\left(-\frac{\nabla^2}{2m} - \mu\right)c_{\sigma,a} - \frac{g}{N
    } \sum_{ab} c^\dagger_{\uparrow,a} c^\dagger_{\downarrow,a}c_{\downarrow,b} c_{\uparrow,b} \right], \label{eq:BCSH}
\end{align}
where $g>0$, and $c_{\sigma,a}(\sigma = \uparrow,\downarrow, a = 1,\ldots,N)$ is a $N$-component spin-$1/2$ fermion.
The partition function can be written down using the coherent state path integral, $Z = \int D(\bar \psi,\psi) \exp(-S_E)$, with the Euclidean action
\begin{align}
    S_E[\bar\psi,\psi] = \int\limits^\beta_0 \ud\tau \int \ud^d x~ \left[\sum_{\sigma,a} \bar\psi_{\sigma,a}\left(\p_\tau - \frac{\nabla^2}{2m} - \mu\right)\psi_{\sigma,a} - \frac{g}{N} \sum_{ab} \bar\psi_{\uparrow ,a}\bar\psi_{\downarrow,a} \psi_{\downarrow,b} \psi_{\uparrow,b}\right], 
\end{align}
where $\bar\psi,\psi$ are (complex conjugate) fermion fields. Using the Hubbard-Stratonovich transformation, the attractive interaction can be written as 
\begin{align}
 \exp\left[\frac{g}{N} \sum_{ab}\int_{\tau,x} \bar\psi_{\uparrow ,a}\bar\psi_{\downarrow,a} \psi_{\downarrow,b} \psi_{\uparrow,b}\right] = \int D(\bar\Delta,\Delta) \exp\left\{ - \int\limits^\beta_0 \ud\tau \int \ud^d x~ \left[\frac{N}{g}\abs{\Delta}^2 - \sum_{a}\left(\bar\Delta \psi_{\downarrow,a} \psi_{\uparrow,a} + \Delta \bar\psi_{\uparrow ,a}\bar\psi_{\downarrow,a}  \right)\right]\right\},
\end{align}
where $\bar\Delta,\Delta$ are the Cooper pair fields, whose equation of motion is $\Delta = \frac{g}{N} \sum_a\psi_{\downarrow,a} \psi_{\uparrow,a} $.
Since the action is now quadratic in the fermion field and takes the form $-\bar\Psi^{\mathrm{T}} \cA \Psi$ where $\Psi = (\psi_\uparrow,\bar\psi_\downarrow)^{\mathrm{T}}$ is the so-called Nambu spinor, we can integrate it out and obtain
\begin{align}
    S_E[\bar\Delta,\Delta] =  N \int\limits_0^\beta\ud\tau \int \ud^d x ~\left[\frac{1}{g}\abs{\Delta}^2 - \left[\tr\log \cA\right]_{xx}\right],
\end{align}
with
\begin{align}
    \cA(x,x') = \begin{pmatrix}
-\p_\tau + \frac{\nabla^2}{2m} +\mu & \Delta \\
\bar\Delta & -\p_\tau - \frac{\nabla^2}{2m} -\mu 
\end{pmatrix} \times \delta(x-x').
\end{align} 
%that is defined in the BdG space. \XYH{Nambu spinor space? BdG probably refer to diagonalizing the mean-field Hamiltonian?} \andy{at least on google ``Nambu spinor" seems to refer to the desired construction}

Now, we generalize the BCS effective action by coupling to an additional $N$-component scalar field $\phi_a$ 
\begin{align}
    S_{\mathrm{eff},E}[\bar\Delta,\Delta,\phi_a,\sigma] = \int\limits^\beta_0 \ud\tau \int \ud^d x~\left[ \frac{1}{2}(\p_\mu \phi_a)^2 + \frac{1}{2}\sigma \left(\phi_a^2-Nt\abs{\Delta}^2\right) + \frac{N}{g} \abs{\Delta}^2   -N \left[\tr\log \cA \right]_{xx}\right] ,
\end{align}
where we extremize over $\sigma$ before evaluating the free energy density, as in the previous model, and $t>0$ is a coupling constant of mass dimension $d-3$.  Before analyzing this model, let us briefly comment on what we have done.  If we attempt to integrate out $\Delta$, we find that Hamiltonian \eqref{eq:BCSH} is coupled to the critical $\mathrm{O}(N)$ model from before with effective four-fermion attractive potential \begin{equation}
    g_{\mathrm{eff}}(\sigma) = \frac{g}{1-\frac{1}{2}gt\sigma }.
\end{equation} 
Recall that $\sigma$ is intuitively related to $\vec\phi^2$.  Evidently, the critical bosons enhance the attraction between fermions which leads to superconductivity. 

Now let us establish the phase diagram of this model.  Suppose $\phi_1  = \phi$ acquires a condensate. Integrating out the rest of $\phi_{a\neq 1}$ we get
\begin{align}
    \frac{1}{N}S_{\mathrm{eff},E}[\bar\Delta,\Delta,\phi,\sigma] = \int\limits_0^\beta\ud\tau \int \ud^d x~\left[ \frac{1}{2}\left[\log(-\p^2+\sigma)\right]_{xx} - \left[\tr \log \cA\right]_{xx} +   \frac{1}{2}(\p_\mu \phi)^2 + \frac{1}{2}\sigma \left(\phi^2-t\abs{\Delta}^2\right) +\frac{1}{g} \abs{\Delta}^2  \right] ,
\end{align}
where we rescaled $\phi\to \sqrt{N}\phi$. Varying with respect to $\phi,\sigma,\abs{\Delta}$, we find
\begin{subequations}\label{eq:SC eom}
\begin{align}
    \sigma \phi &= 0 \label{eq:phi eom}\\
    T\sum_n \int \frac{\ud^d k}{(2\pi)^d} \frac{1}{\omega_n^2 + k^2 +\sigma} + \left(\phi^2-t\abs{\Delta}^2\right)& = 0 \label{eq:sigma eom}\\
    \left(\frac{1}{g}-\sigma t - 2T \nu \int_0^{\omega_*}\ud \xi~ \sum_n \frac{1}{\Omega_n^2 + \xi^2+\abs{\Delta}^2} \right)\abs{\Delta} & = 0, \label{eq:Delta eom}
\end{align}
\end{subequations}
where $\omega_n = 2n \pi T$ and $\Omega_n = (2n+1) \pi T$, and we have approximated $\int \ud^d k/(2\pi)^d \approx \nu \int_{-\omega_*}^{\omega_*}\ud \xi$ with $\nu$ the density of state at the Fermi level which is determined by $\xi(k)\equiv k^2/2m - \mu = 0 $.  When the Fermi surface is finite, $\nu$ is essentially a constant parameter.  We also expect that the subsequent calculation is largely insensitive to the microscopic shape of the Fermi surface or its dispersion relation. In the above equation, we have introduced a UV-cutoff $\omega_*$, the Debye frequency, so that we are interested in $\abs{\Delta},T\ll \omega_*$. To simply determine the orders, it is convenient to ignore the phase of the order parameter and choose $\Delta = \abs{\Delta}$.

The usual BCS gap equation is obtained by setting $t = 0$ in \eqref{eq:Delta eom}, which leads to
\begin{align}\label{eq:gap eq}
    \frac{1}{g \nu} = \int\limits_0^{\beta \omega_*} \ud x ~ \frac{\tanh\left(\frac{1}{2}\sqrt{x^2 + (\beta\Delta)^2}\right)}{\sqrt{x^2 + (\beta\Delta)^2}}.
\end{align}
Since the integral is monotonically decreasing when $\Delta$ is increasing, its maximum value at a fixed $T$ is achieved by setting $\Delta=0$. If this is still smaller than $1/g\nu$, we find no solutions for $\Delta\neq 0$. The critical temperature above which there exists no solutions is given by
\begin{align}
    T_{\mathrm{c}} \sim  \omega_* \exp(-1/g\nu),
\end{align}
which is known as the BCS critical temperature. When $2<d<3$, $t$ is irrelevant from its mass dimension, but, as we show below, the solutions change dramatically as soon as $t>0$. We therefore regard $t$ as a ``dangerously irrelevant'' coupling.

Now, let us solve \eqref{eq:SC eom} for $\Delta\neq 0$. Performing the Matsubara summation, we find 
\begin{subequations}
    \begin{align}
        \int \frac{\ud^d k}{(2\pi)^d} \frac{1}{\sqrt{k^2+\sigma}(e^{\beta\sqrt{k^2+\sigma}}-1)} + \phi^2 -t\Delta^2&= 0 \label{eq:sc eq 1} \\
        \frac{1}{g}-\sigma t - \nu \int\limits_0^{\beta \omega_*} \ud x ~ \frac{\tanh\left(\frac{1}{2}\sqrt{x^2 + (\beta\Delta)^2}\right)}{\sqrt{x^2 + (\beta\Delta)^2}}& =0, \label{eq:sc eq 2}
    \end{align}
\end{subequations}
where we have the propagator at coinciding points by proper renormalization of the bare mass in the zero temperature QFT.
Since $\sigma\phi=0$, we look at three sets of solutions $(\sigma = 0,\phi\neq 0)$ or $(\sigma \neq 0,\phi =  0)$ or $(\sigma=\phi=0)$. In the zero temperature limit, $\beta\to \infty$, \eqref{eq:sc eq 1} is approximately given by $\phi^2 \approx t \Delta^2$. Hence, to support $\Delta^2\neq 0$, we must set $\sigma=0$. From \eqref{eq:sc eq 2}, we then have 
\begin{align}
    \Delta(T\to 0) \sim \omega_*\exp(-1/g\nu),
\end{align}
which corresponds to the usual BCS gap. At low $T$, we have both superconductivity and $\mathrm{O}(N)\to \mathrm{O}(N-1)$ condensation. Upon increasing $T$ and keeping $\sigma=0$, the BCS gap $\Delta$ will decrease since the $x$-integral in \eqref{eq:sc eq 2} is a decreasing function of $T$. This results in a decrease of $\phi^2$ according to \eqref{eq:sc eq 1}. But since $\phi^2$ cannot be negative, there must be a phase transition at some temperature $T_{\mathrm{c}}^\prime$.
%whether this transition is continuous, first order, or separated by an intermediate phase, e.g. when $T_{\mathrm{c}}<T_{\mathrm{c}}^\prime$, goes beyond our scope. 
Assuming a continuous transition, $T_{\mathrm{c}}^\prime$ can be determined as follows. At $\sigma = \phi^2=0$, \eqref{eq:sc eq 1} is solved by, at $d>2$,
\begin{align}
    \Delta^2(T_{\mathrm{c}}^\prime) \approx
    t^{-1}\int \frac{\ud^d k}{(2\pi)^d} \frac{1}{k(\mathrm{e}^{k/T_{\mathrm{c}}^\prime}-1)} = c_d t^{-1} T_{\mathrm{c}}^{\prime d-1},
\end{align}
where $c_d$ is a constant. Then, \eqref{eq:sc eq 1} implies
\begin{equation}\label{eq:sc tc prime}
\frac{1}{g} = \nu \int\limits_0^{\beta_c'\omega_*}\mathrm{d}x  \frac{\tanh\left(\frac{1}{2}\sqrt{x^2 + (\beta_c'\Delta(T_{\mathrm{c}}^\prime))^2}\right)}{\sqrt{x^2 + (\beta_c'\Delta(T_{\mathrm{c}}^\prime))^2}}  \approx \nu \log \frac{\omega_*}{\Delta(T_{\mathrm{c}}^\prime)} = \nu \log \frac{\omega_*}{\sqrt{c_d}t^{-1/2}(T_{\mathrm{c}}^\prime)^{(d-1)/2}}.
\end{equation}
In the high temperature phase $T>T_{\mathrm{c}}^\prime$, we have $\sigma > 0$, and if we assume that $\sigma\ll T^2$, \eqref{eq:sc eq 1} would imply
\begin{align}\label{eq:high T Delta}
    \Delta^2(T>T_{\mathrm{c}}^\prime) \sim t^{-1} T^{ d-1}
\end{align}
which indicates that $\Delta^2\neq 0$ persists to any high $T$. Then, \eqref{eq:sc eq 2} allows us to estimate
\begin{align}
    \sigma \approx \frac{1}{g t} + \frac{\nu}{t} \log\left(\frac{\Delta}{\omega_*}\right) \sim \frac{1}{g t} + \frac{\nu}{t} \log\left(\frac{T^{(d-1)/2}}{t^{1/2}\omega_*}\right),
\end{align}
which indeed satisfies $\sigma\ll T^2$. Our assumption is sensible if $(gt)^{-1/2} \ll T_{\mathrm{c}}^\prime $, and using \eqref{eq:sc tc prime}, we find
\begin{align}
    T_{\mathrm{c}}^\prime\sim \left(\omega_* \mathrm{e}^{-1/g\nu} t^{1/2}\right)^{2/(d-1)} \gg (gt)^{-1/2}.
\end{align}
It is clear that there would be no other solutions particularly with $\Delta = 0$, since $\sigma$ would be too small to satisfy \eqref{eq:sc eq 1} provided the above assumptions hold. Therefore, at high $T$, we have superconductivity without scalar condensation, and the order would persist to high temperature.

\bibliography{thebib}

\end{document}